\documentclass[sigconf]{acmart}

\AtBeginDocument{%
  }

\usepackage{multirow}
\usepackage{algorithm}
\usepackage{algorithmic}
\usepackage{comment}
\usepackage{color}
\usepackage{subfigure}
\usepackage{enumitem}

\copyrightyear{2023}
\acmYear{2023}
\setcopyright{acmlicensed}\acmConference[SIGIR '23]{Proceedings of the 46th
International ACM SIGIR Conference on Research and Development in
Information Retrieval}{July 23--27, 2023}{Taipei, Taiwan}
\acmBooktitle{Proceedings of the 46th International ACM SIGIR Conference on
Research and Development in Information Retrieval (SIGIR '23), July 23--27,
2023, Taipei, Taiwan}
\acmPrice{15.00}

\begin{document}

\title{Law Article-Enhanced Legal Case Matching: \\a Causal Learning Approach}

\author{Zhongxiang Sun}
\affiliation{
  \institution{Gaoling School of Artificial Intelligence\\Renmin University of China}
\city{Beijing}
  \country{China}
  }
\email{sunzhongxiang@ruc.edu.cn}

\author{Jun Xu}
\authornote{Jun Xu is the corresponding author. Work partially done at Engineering Research Center of Next-Generation Intelligent Search and Recommendation, Ministry of Education.}
\affiliation{
  \institution{Gaoling School of Artificial Intelligence\\Renmin University of China}
    \city{Beijing}
  \country{China}
  }
\email{junxu@ruc.edu.cn}

\author{Xiao Zhang}
\affiliation{
  \institution{Gaoling School of Artificial Intelligence\\Renmin University of China}
  \city{Beijing}
  \country{China}
  }
\email{zhangx89@ruc.edu.cn}

\author{Zhenhua Dong}
\affiliation{%
  \institution{Noah's Ark Lab, Huawei}
  \city{Shenzhen}
  \country{China}
  }
\email{dongzhenhua@huawei.com}

\author{Ji-Rong Wen}
\affiliation{%
  \institution{Gaoling School of Artificial Intelligence\\Renmin University of China}
  \city{Beijing}
  \country{China}
  }
\email{jrwen@ruc.edu.cn}

\renewcommand{\shortauthors}{Zhongxiang Sun, Jun Xu, Xiao Zhang, Zhenhua Dong, \& Ji-Rong Wen} 

\begin{abstract}
Legal case matching, which automatically constructs a model to estimate the similarities between the source and target cases, has played an essential role in intelligent legal systems.
Semantic text matching models have been applied to the task where the source and target legal cases are considered as long-form text documents. These general-purpose matching models make the predictions solely based on the texts in the legal cases, overlooking the essential role of the law articles in legal case matching. In the real world, the matching results (e.g., relevance labels) are dramatically affected by the law articles because the contents and the judgments of a legal case are radically formed on the basis of law. From the causal sense, a matching decision is affected by the mediation effect from the cited law articles by the legal cases, and the direct effect of the key circumstances (e.g., detailed fact descriptions) in the legal cases. In light of the observation, this paper proposes a model-agnostic causal learning framework called Law-Match, under which the legal case matching models are learned by respecting the corresponding law articles. Given a pair of legal cases and the related law articles, Law-Match considers the embeddings of the law articles as \emph{instrumental variables} (IVs), and the embeddings of legal cases as \emph{treatments}. Using IV regression, the treatments can be decomposed into law-related and law-unrelated parts, respectively reflecting the mediation and direct effects. These two parts are then combined with different weights to collectively support the final matching prediction. We show that the framework is model-agnostic, and a number of legal case matching models can be applied as the underlying models. Comprehensive experiments show that Law-Match can outperform state-of-the-art baselines on three public datasets.
\end{abstract}



\maketitle
\section{Introduction}
Legal case matching has played an important role in intelligent legal systems. For example, in legal case retrieval, the matching models help the system to determine the relevance between the query cases and the candidate cases. Traditionally, the task is formalized as matching two long-form text documents at the semantic level. General-purpose document matching models have been adapted to tackle the problem, including the
heuristic methods~\cite{saravanan2009improving,zeng2005knowledge}, network-based methods~\cite{bhattacharya2020hier,bhattacharya2020methods}, and text-based methods~\cite{shao2020bert,xiao2021lawformer}.

Though effective, 
simply considering the legal cases as general long-form text documents~\cite{yu-etal-2022-optimal} still has spaces for improvement. One striking difference between legal cases and general documents is that legal cases usually cite a number of law articles\footnote{Law articles are the foundation of statutes or written laws which are usually enacted by the administration of justice (e.g., Criminal Law in China).}. 
These law articles are selected from the law book (e.g., Chinese Criminal Law) by the judges and provide essential knowledge of the legal case's context and judgments. Existing studies have shown that law articles are beneficial to a number of legal-related tasks~\cite{zhong2018legal, xu2020distinguish, sun2023short}.

\begin{figure}[t]
    \centering
    \includegraphics[width=\linewidth]{ 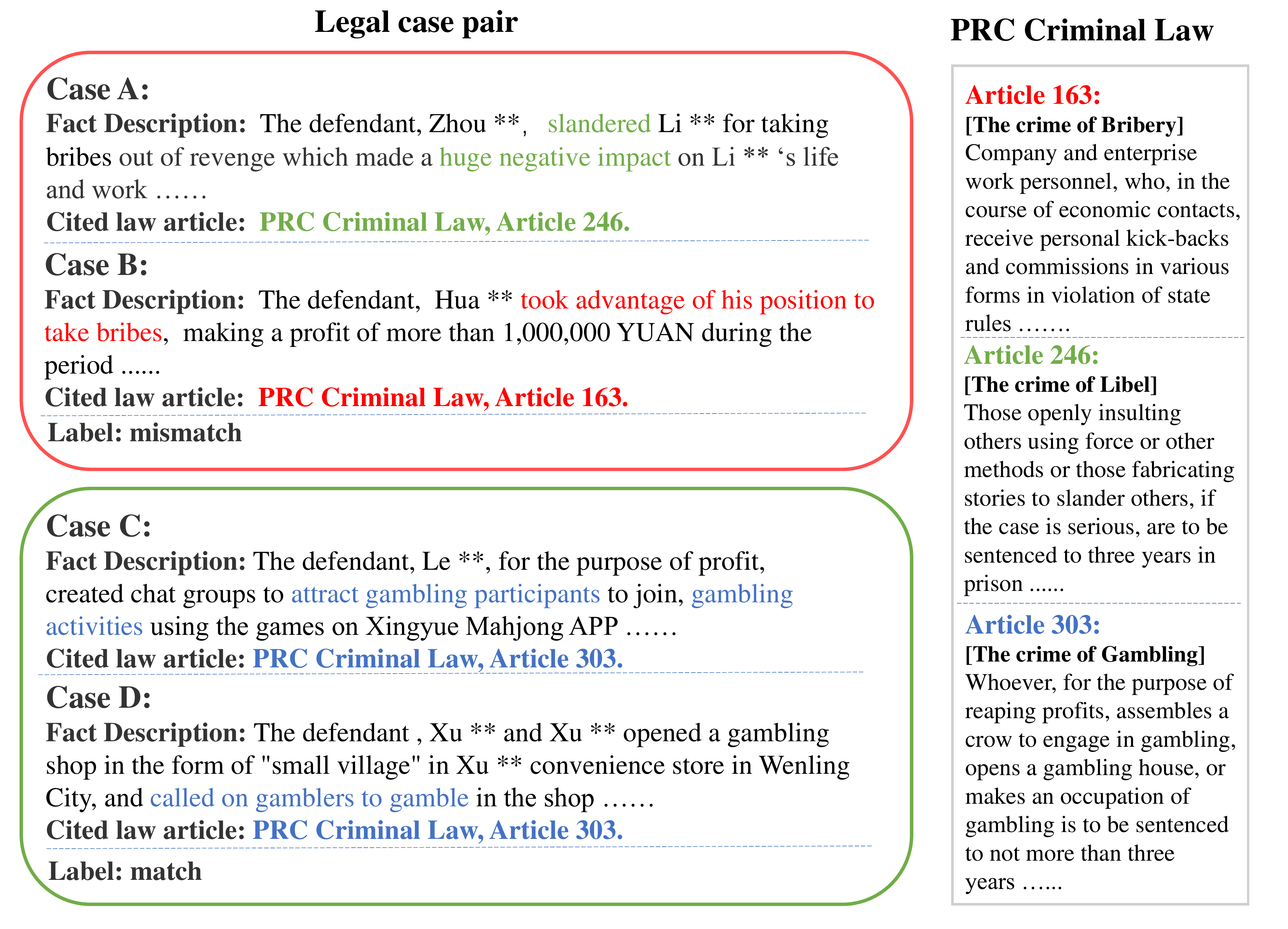}
    \caption{Left: two pairs of legal cases; Right: three cited law articles in the legal cases. (translated from Chinese)}
    \label{fig:intro}
\end{figure}

Analysis shows that the law articles are also important in legal case matching.
\autoref{fig:intro} shows the snapshots of two real legal case pairs\footnote{Crawled from \url{http://faxin.cn} and translated to English.}. Contents of the cited law articles are listed in the right part of the figure. In the first legal case pair, Case A and Case B share a large number of words in their fact descriptions. However, the judges' decisions are: Case A is libel crime (PRC Criminal Law, Article 246) while Case B is bribery crime (Article 163). The associated law articles are helpful in identifying the key information (highlighted) in the two cases~\cite{feng-etal-2022-legal}. By comparing the key information in the two cases, experts annotate the matching label as ``mismatch'' though they have relatively high semantic text similarity (measured by Lawformer~\cite{xiao2021lawformer}). 
In the second example of \autoref{fig:intro}, Case C and Case D have relatively lower semantic text similarity than the previous pair. However, both of them are judged as the gambling crime (Article 303). The law article helps to identify similar key information (highlighted) in these two legal cases. So the expert-annotated matching label is  ``match''.



Usually, the key constitutive elements and the key circumstances provide important signals for the matching of two legal cases~\cite{ma2021lecard}. 
The key constitutive elements are highly summative texts written in light of some law articles. The key circumstances, on the other hand, are detailed fact descriptions and are usually very different from case to case. They are not directly related to any law articles. Therefore, it is possible that law articles can help the matching model to identify and decompose these key information.

From the causal sense, the matching of two legal cases is affected by the \emph{mediation effect} from the law articles and the \emph{direct effect} from the key circumstances part of legal cases. 
More specifically, the key constitutive elements in the legal cases mediate the law articles' effect on the matching decision (i.e., the mediation effect). In contrast, the key circumstances directly affect the matching decision (i.e., the direct effect). As a result, the embedding of a legal case actually consists of two parts: the law-related part, which is the mediator of the mediation effect, and the law-unrelated part, which has direct effect. These two parts reflect different association mechanisms between the legal cases and the matching decisions. It is necessary to identify and treat them differently.



To address the issue, this paper proposes a causal representation learning framework tailored for legal case matching, called Law-Match. Specifically, Law-Match considers the legal cases as \emph{treatment} and the corresponding law articles as \emph{instrument variables} (IVs)~\cite{angrist1996identification,caner2004instrumental,stock2003retrospectives,venkatraman2016online}. 
In the matching phase, after getting the embeddings of the legal cases (i.e., treatments) and the related law articles (i.e., IVs), Law-Match first uses the IVs to regress the treatments, resulting in the fitted vector (law-related part) and the residuals (law-unrelated part). These two parts have different effects on the final matching. Law-Match then combines them into a newly reconstructed treatment vector with the attention mechanism. Finally, the reconstructed treatment is fed to the underlying matching model for making the final matching prediction. In the training phase, 
an alternative optimization procedure is developed to learn the parameters in the IV regression and matching models.  

We summarize the major contributions of the paper as follows:

(1) We analyze the essential role of law articles in legal case matching from a causal view: the matching decisions are affected by the mediation effect of the law articles and the direct effect of the key circumstances in the legal cases. 

(2) We propose a novel model-agnostic causal learning framework which introduces the law articles into the process of legal case matching in a theoretically sound way. IV regression is adopted to decompose the mediation effect and direct effect from the legal case embeddings by considering law articles as IVs and legal cases as the treatments. 

(3) We conducted extensive experiments on three public datasets. Experimental results demonstrated that Law-Match could significantly improve the underlying models' performance and outperform the baselines, verifying the importance of the law articles in legal case matching.

\section{Related Work}
\subsection{Legal case matching}
Conventionally, legal case matching can be addressed with manual knowledge engineering (KE)~\cite{2012A}. The methods include the Boolean search technology and manual classification~\cite{dias2022state}. With the development of NLP, deep learning has been adapted to realize semantic level matching of legal cases. According to~\cite{bhattacharya2020methods}, these studies can be categorized as network-based and text-based methods. The network-based methods are tailored for common law and use the citations of different cases to construct a Precedent Citation Network (PCNet). For example,~\cite{kumar2011similarity} use PCNet-based Jaccard similarity to infer the paired legal cases' similarity.~\citet{bhattacharya2020methods} use Node2vec to map the nodes of the graph to a vector space and then compute the legal cases' cosine similarity. See also~\cite{minocha2015finding, bhattacharya2020hier}. 

The text-based methods compute the semantic similarity between legal cases.
~\citet{shao2020bert} utilize BERT to capture the semantic relationships at the paragraph level and then infer the relevance between two cases by aggregating the paragraph-level interactions.~\citet{xiao2021lawformer} release the longformer-based~\cite{beltagy2020longformer}  pre-trained language model to get a better representation of long legal documents.~\citet{yu2022explainable} propose a three-stage explainable legal case matching model. 
 Law articles have shown their effects on a number of legal tasks. \citet{zhong2018legal} jointly model the law article prediction task and the Legal Judgment Prediction (LJP).~\citet{xu2020distinguish} construct a relationship diagram between the law articles and introduce all relevant law articles into the LJP.

\subsection{Causal learning}
In causal learning, \textbf{instrument variable (IV)}~\cite{angrist1996identification,caner2004instrumental,stock2003retrospectives,venkatraman2016online} has been widely used to identify the causal effect of the treatment on the output.
Traditionally, two-stage least squares (2SLS)~\cite{angrist1995two} regression is used to regress the IVs to treatments.~\citet{shaver2005testing}  and~\citet{dippel2020causal} use the linear 2SLS to identify both the causal treatment and mediation effects. Recently, models have been proposed to extend the linear 2SLS model to high dimensional and non-linear deep neural networks. 
~\citet{xu2020learning} extend 2SLS to an alternating training regime and perform well in high-dimensional image data and off-policy reinforcement learning. See also~\cite{si_enhancing,hartford2017deep,niu2022estimation,yuan2022auto,wu2022treatment}. 


Recently, \textbf{causal representation learning} has been proposed to discover the high-level causal variables from low-level observations~\cite{scholkopf2021toward}.~\citet{yang2021causalvae} propose a VAE-based causal disentangled representation learning framework by leveraging labels of concepts as additional knowledge.~\citet{si2022model} reconstruct the causal representation of items in recommendation by using search data as the additional knowledge. 
The proposed Law-Match can also be viewed as learning causal legal case representations by using the law articles as additional knowledge. 

\textbf{Mediation analysis} is designed to explore the underlying mechanism by which one variable influences another variable through a mediator variable~\cite{dodhia2005review}. 
In order to better leverage the different mechanisms, many studies are proposed to decompose the different effects~\cite{vanderweele2013three,bellavia2018decomposition}. In this paper, we also use the IV regression to identify the indirect effects between law articles and matching results.



\section{Background and Preliminaries}

\subsection{Problem formulation}
Suppose we are given a set of labelled data tuples $\mathcal{D} = \{(X, Y, z)\}$ where $X\in \mathcal{X}$ is a source legal case, $Y\in \mathcal{Y}$ is a target case, and $z\in\mathcal{Z}$ is the human-annotated matching label. $\mathcal{Z}$ could be defined as, for example, $\mathcal{Z}=\{0, 1, 2\}$ where 0 means mismatch, 1 means partially match, and 2 means match. Typically, a legal case can be considered as a sequence of words that describe the case's facts. 
Therefore, the legal case $X$ (or $Y$) can be represented as a $d$-dimensional embeddings $\mathbf{e}(X)\in\mathbb{R}^d$ (or $\mathbf{e}(Y)\in\mathbb{R}^d$).
Typically, the embeddings are the outputs at the [CLS] token of a BERT model pre-trained on a legal corpus.

The task of legal case matching, therefore, becomes learning a matching model $f:\mathcal{X}\times\mathcal{Y}\rightarrow \mathcal{Z}$ based on the labelled tuples in $\mathcal{D}$. 

\subsection{Law articles in legal case matching}
Real legal cases usually cite the applicable law articles that support the judicial decisions\footnote{In some tasks such as legal retrieval, the source cases (queries) only contain the fact descriptions. For these cases, the law articles can be extracted with the causal discovery method~\cite{liueverything}. Please refer to Section~\ref{subsec:MissLawArticle} for more details.}.
These law articles are selected from a law book (e.g., PRC Criminal Law). The IDs (i.e., IDs of the articles, clauses, and items) are listed at the end of legal cases. Therefore, we can collect the article contents from the law book according to the IDs. The law article contents can be concatenated as a new pseudo document, represented as another $d$-dimensional embedding $\mathbf{e}(L_X)\in\mathbb{R}^d$ where $L_X$ denotes the law articles cited by $X$. 

Intuitively, the law articles should provide complementary knowledge for understanding the legal cases and therefore enhancing the matching. On the one hand, legal cases are long-form documents containing multiple sentences, describing a number of facts. Some of them are the key facts, while others are not. The applicable law articles are selected by the judges. They should reflect the most key information (e.g., key facts) in the case, affecting the legal case matching. On the other hand, the law articles influence the description of the facts and the judgments (e.g., charges, terms of penalty). When preparing a legal case, the lawyers would consider the law articles seriously because the judge's decisions are based on the law articles. 
 

One straightforward approach is concatenating the contents of the law articles to the original texts, i.e., appending $L_X$ to $X$ and appending $L_Y$ to $Y$. Though improvements can be observed, we note that there exist fundamental differences between the law articles and texts in legal cases: the law articles are created by the governmental institutions and presented in the form of general rules with precise definitions. The legal cases are written by judges in the form of detailed descriptions of specific facts. They have different roles and affect the matching with different mechanisms. 


\subsection{A causal view of legal case matching}\label{sec: view in causal}

\begin{figure}[tp]
    \subfigure[Causal graph of legal case matching.]{
    \includegraphics[width=0.45\linewidth]{ 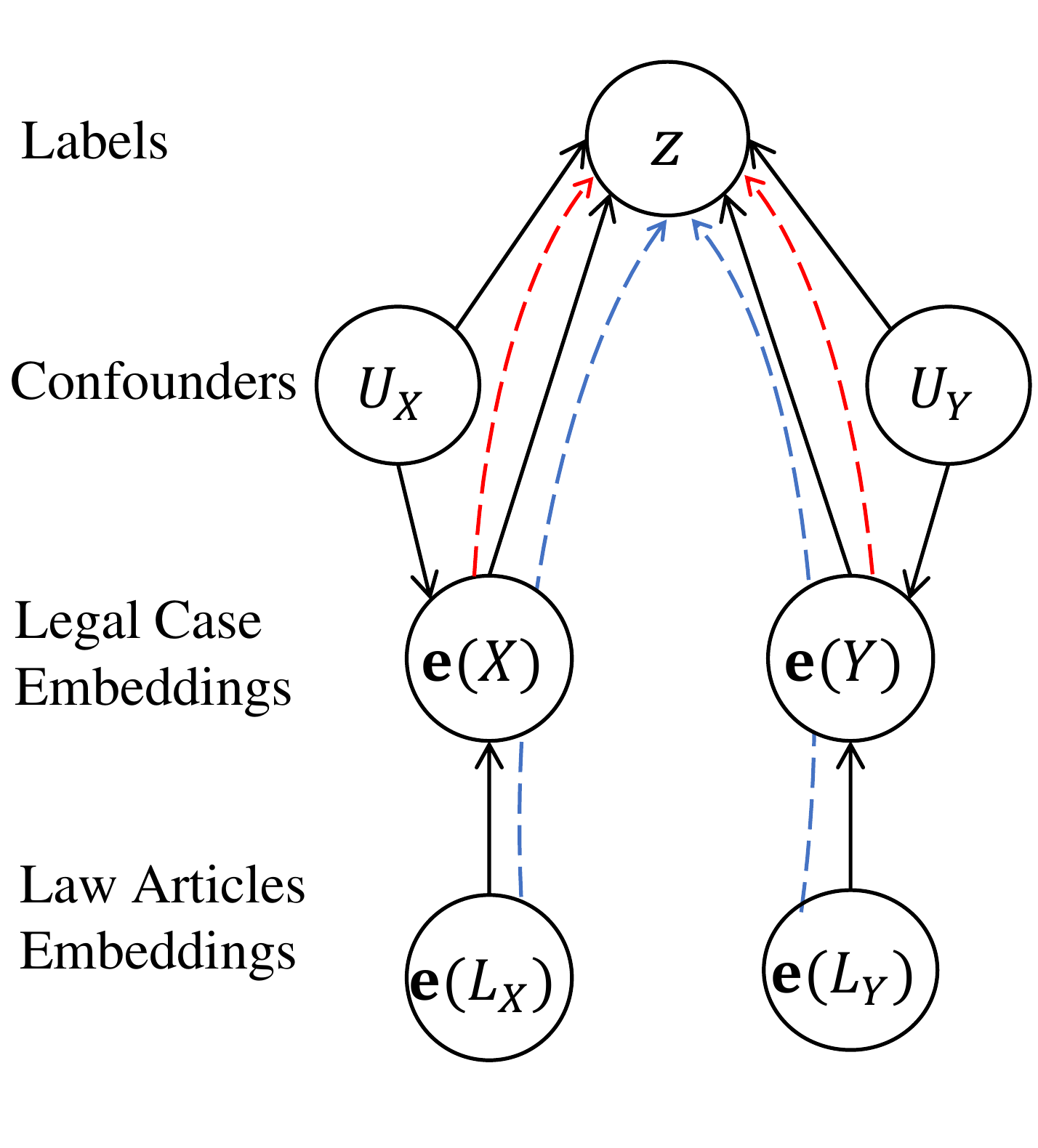}
    \label{fig:origin_causal_graph}
    }
    \subfigure[Causal graph after IV regression.]{
    \includegraphics[width=0.495 \linewidth]{ 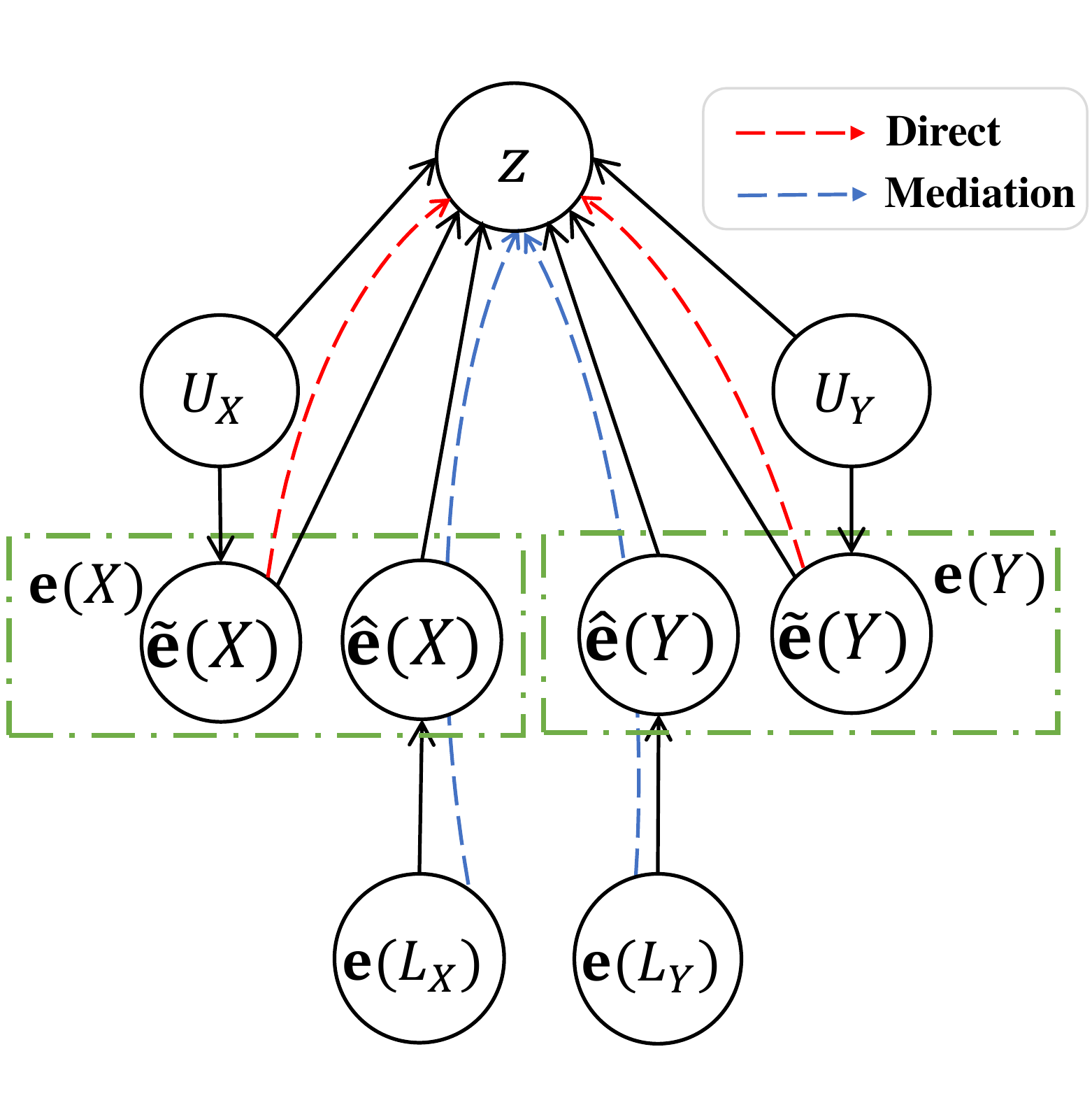}
    \label{fig:IV_causal_graph}
    }
    \caption{Causal graph of legal case matching and the graph after IV regression.
    (a): Law article embeddings $\mathbf{e}(L_{X})$ and $\mathbf{e}(L_{Y})$ affect $z$ through $\mathbf{e}(X)$ and $\mathbf{e}(Y)$, which also have other effects on $z$.
    (b): $\mathbf{e}(L_{X})$ and $\mathbf{e}(L_{Y})$ affect $z$ through the fitted parts (mediators) $\widehat{\mathbf{e}}(X)$ and $\widehat{\mathbf{e}}(Y)$. The residuals $\widetilde{\mathbf{e}}(X)$ and $\widetilde{\mathbf{e}}(Y)$ have direct effects on $z$.}
    \label{fig:Causal_graphs}
\end{figure}

    


Following the framework proposed in~\cite{pearl2009causality, peters2017elements}, we can formalize legal case matching with a multivariate causal graph. According to Figure~\ref{fig:origin_causal_graph}, the two input legal cases $X$ and $Y$ are \emph{two treatment variables} in the causal graph, respectively represented as their embeddings $\mathbf{e}(X)$ and $\mathbf{e}(Y)$. The \emph{outcome variable} $z$ is the matching label. There exist associations between $\mathbf{e}(X)$ and $z$ (path $\mathbf{e}(X)\rightarrow z$) and $\mathbf{e}(Y)$ and $z$ (path $\mathbf{e}(Y)\rightarrow z$), because the prediction is based on the matching signals between $X$ and $Y$. 

The observations in~\citet{ma2021lecard} show that the key constitutive elements in a legal case are generally highly related to the cited law articles. The key circumstances, however, are not. 
In a causal sense, the matching labels are determined along two paths, including the \emph{mediation effect} of law articles and the \emph{direct effect} of the key circumstances. 
More specifically, \emph{the key constitutive elements mediate the effect of the law articles on the matching label}, while the key circumstances have direct effects on the matching label. 
Therefore, the associations $\mathbf{e}(X)\rightarrow z$ and $\mathbf{e}(Y)\rightarrow z$ are mixtures of two different types of causal paths, i.e., the law-related associations caused by the mediation effect and 
the law-unrelated associations caused by the direct effect. 

Besides, for legal case $X$ (or $Y$), there also exist missing variables $U_X$ (or $U_Y$) that are associated with both $X$ and $z$ (paths $\mathbf{e}(X)\leftarrow U_X\rightarrow z$). 
The missing variables could be any confounding factors unrelated to law articles (e.g., the focus of disputes). However, they are important parts of the legal case and are considered when making the matching decisions. Therefore, the law-unrelated association can be viewed as a backdoor path in the causal graph. 
Obviously, the law-related and law-unrelated associations reflect different mechanisms between legal cases (i.e., treatments) and matching prediction (i.e., outcome). 
It is necessary to identify these two associations and treat them differently.

The independence between the law articles and the missing variables as well as the key circumstances, provide us a chance to conduct the identification. 
As shown in Figure~\ref{fig:IV_causal_graph}, we leverage $L_X$ and $L_Y$ as the IVs~\cite{xu2020learning,wooldridge2015introductory}\footnote{According to~\citet{wooldridge2015introductory}, $\widetilde{\mathbf{e}}(X)$ will contain an error term if we only use ${\mathbf{e}(L_X)}$ to regress $\mathbf{e}(X)$. We cannot control the association between $\widehat{\mathbf{e}}(X)$ and $\widetilde{\mathbf{e}}(X)$.}. 
Thus, we can regress $\mathbf{e}(X)$ on $\mathbf{e}(L_X)$ and $\mathbf{e}(L_Y)$ to get $\widehat{\mathbf{e}}(X)$ which does not depend on the confounder $U_X$ and residual $\widetilde{\mathbf{e}}(X)=\mathbf{e}(X) - \widehat{\mathbf{e}}(X)$. Therefore, path $\widehat{\mathbf{e}}(X)\rightarrow z$ can be viewed as purely law-related associations. Paths $\widetilde{\mathbf{e}}(X)\rightarrow z$ and $\widetilde{\mathbf{e}}(X)\leftarrow U_X\rightarrow z$ can be seen as totally law-unrelated. Similarly, we can regress the embedding of $\mathbf{e}(Y)$ on 
$\mathbf{e}(L_X)$ and $\mathbf{e}(L_Y)$, to get $\widehat{\mathbf{e}}(Y)$ which does not depend on the confounder $U_Y$, and the residual part $\widetilde{\mathbf{e}}(Y) =\mathbf{e}(Y)- \widehat{\mathbf{e}}(Y)$. $\widehat{\mathbf{e}}(Y)\rightarrow z$ can be viewed as law-related associations. $\widetilde{\mathbf{e}}(Y)\rightarrow z$ and $\widetilde{\mathbf{e}}(Y)\leftarrow U_Y\rightarrow z$ can be seen as law-unrelated associations. In this way, we can identify the law-related associations and law-unrelated associations under a causal framework. 

 \section{Our Approach: Law-Match}
\label{sec: main model}
This section presents an implementation of the causal learning framework for legal case matching, called Law-Match.

\begin{figure}[tp]
    \centering
    \includegraphics[width=\linewidth]{ 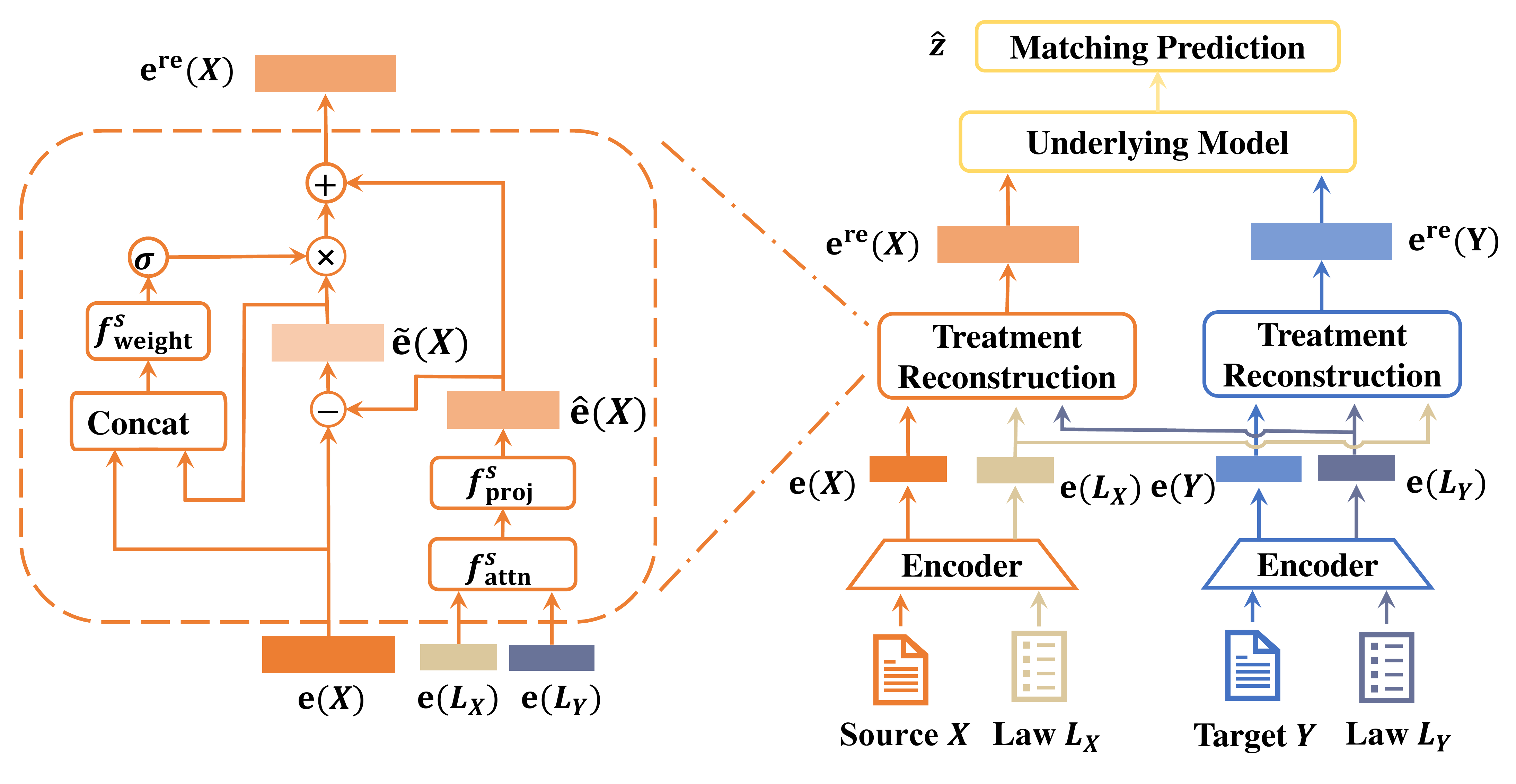}
    \caption{The architecture of Law-Match. Right: procedure of Law-Match applied to an underlying text matching model; Left: procedure of treatment reconstruction. }
    \label{fig:architecture}
\end{figure}

\subsection{Model overview}
\autoref{fig:architecture} illustrates the architecture of Law-Match. Given a pair of legal cases $(X, Y)$, Law-Match first encodes them as two embeddings (two treatments). Also, the cited law articles $L_X$ and $L_Y$ are encoded as two embeddings (two IVs). Then, the treatment reconstruction module is employed to decompose each treatment into two vectors with the help of the corresponding IVs. After that, the decomposed vectors are combined as a new reconstructed vector. Finally, the reconstructed treatment vectors are fed to the downstream matching model for making the matching prediction. 



\subsection{Treatment reconstruction}
As shown in the left part of~\autoref{fig:architecture}, Law-Match employs two treatment reconstruction modules to process the source case $X$ and target case $Y$, outputs the reconstructed embeddings $\mathbf{e}^\mathrm{re}(X)$ and $\mathbf{e}^\mathrm{re}(Y)$, respectively. These two modules share the same network architecture while with different parameters. 


As have shown in Section~\ref{sec: view in causal}, the new treatment $\mathbf{e}^\textrm{re}(X)$ can be created by first regressing $\mathbf{e}(X)$ on the IVs $\mathbf{e}(L_X)$ and $\mathbf{e}(L_Y)$, achieving the fitted part and residual part. We call this stage treatment decomposition. Then, these two parts are re-combined together with attended weights, called treatment reconstruction. 

\subsubsection{Treatment decomposition}
IV regression is used to decompose $\mathbf{e}(X)$ into the fitted part $\widehat{\mathbf{e}}(X)$ and residual part $\widetilde{\mathbf{e}}(X)$, with the help of IVs $\mathbf{e}(L_X)$ and $\mathbf{e}(L_Y)$. Specifically, $\widehat{\mathbf{e}}(X)$ can be written as
\begin{equation}
\label{eqn:proj}
\widehat{\mathbf{e}}(X) = f_{\mathrm{proj}}^{s}\left(c_s(L_X, L_Y)\right),
\end{equation}
where $f_{\mathrm{proj}}^{s}:\mathbb{R}^d\rightarrow\mathbb{R}^d$ is a projection network that maps a law article embedding to the space of legal case embeddings, and the input vector $c_s(L_X, L_Y)$ is a linear combination of the two law article embeddings $\mathbf{e}(L_X)$ and $\mathbf{e}(L_Y)$:
\[
c_s (L_X, L_Y) = w_s\cdot \mathbf{e}(L_X) + (1-w_s)\cdot \mathbf{e}(L_Y),
\]
where
\[
w_s = \frac{\exp\{f_{\mathrm{attn}}^{s}\left(\mathbf{e}(X),\mathbf{e}(L_X)\right)\}}{\exp\{f_{\mathrm{attn}}^{s}\left(\mathbf{e}(X),\mathbf{e}(L_X))\right\}+\exp\{f_{\mathrm{attn}}^{s}\left( \mathbf{e}(X),\mathbf{e}(L_Y)\right)\}},
\]
and $f_{\mathrm{attn}}^{s}(\cdot, \cdot)$ denotes the additive attention~\cite{bahdanau2014neural}:
\[
f_{\mathrm{attn}}^{s}(\mathbf{a},\mathbf{b}) =  \mathbf{v}^{\mathrm{T}}\mathrm{tanh}(\mathbf{W}[\mathbf{a};\mathbf{b}]),
\]
where $\mathbf{v}$ and $\mathbf{W}$ are learnable attention parameters and `$[\cdot;\cdot]$' denotes concatenation of two vectors. The fitted part $\widehat{\mathbf{e}}(X)$ reflects the law-related median association between the law article embeddings and the matching results. 

Given $\widehat{\mathbf{e}}(X)$, it is easy to get the residual
part:
\begin{equation}
\label{eqn:e_tilde}
\widetilde{\mathbf{e}}(X) = \mathbf{e}(X) - \widehat{\mathbf{e}}(X).
\end{equation} 
Obviously, $\widetilde{\mathbf{e}}(X)$ reflects the law-unrelated direct association between the legal case embeddings and the matching results.

\subsubsection{Treatments reconstruction}
The fitted parts and the residuals can be recombined, achieving a new treatment:
\begin{equation}
\label{eqn:recon}
\mathbf{e}^{\mathrm{re}} (X) = \widehat{\mathbf{e}}(X)+ \alpha_{s}\cdot \widetilde{\mathbf{e}}(X),
\end{equation}
where $\alpha_{s}\in[0,1]$ re-weights the influence of the residual part:
\[
\alpha_{s} =\sigma(f_{\mathrm{weight}}^{s}([\mathbf{e}(X); \widetilde{\mathbf{e}}(X)])),
\]
where $f_{\mathrm{weight}}^{s}$ denotes a two-layer MLP that takes the concatenation of the treatments and the residual part as input and outputs a real number, $\sigma$ denotes the sigmoid function. 

Similarly, given the embedding of the target legal case $\mathbf{e}(Y)$ and law article embeddings $\mathbf{e}(L_Y)$ and $\mathbf{e}(L_X)$, we can also get the reconstructed treatment through IV regression and recombination:
\begin{equation}
\label{eqn:proj_Y}
\begin{split}
    \widehat{\mathbf{e}}(Y) =& f_\mathrm{proj}^t\left(c_t (L_Y, L_X)\right),\\
   \widetilde{\mathbf{e}}(Y) =&   {\mathbf{e}(Y)} -\widehat{\mathbf{e}}(Y) ,\\
   \mathbf{e}^\mathrm{re} (Y) = & \widehat{\mathbf{e}}(Y)+ \alpha_{t}\cdot \widetilde{\mathbf{e}}(Y),
\end{split}
   \end{equation}
where $f_\mathrm{proj}^t$, $c_t$, and $\alpha_{t}$ are defined similarly as their counterparts (i.e., $f_\mathrm{proj}^s$ and  $c_s$ in Equation~(\ref{eqn:proj}), and $\alpha_{s}$ in Equation~(\ref{eqn:recon})).

\subsection{Model-agnostic application}
Many document matching models share a similar structure, which we refer to as the underlying model. The underlying models represent each input document as an embedding vector and predict the matching score based on the representations.
Law-Match is a model-agnostic framework implemented over existing document matching models that follow this underlying structure by feeding the reconstructed treatments to the matching model. 

Formally, given a pair of legal cases $(X, Y)$, the two treatment reconstruction modules respectively output the reconstructed embeddings $\mathbf{e}^\textrm{re} (X)$ and $\mathbf{e}^\textrm{re} (Y)$. Then, the matching score between $X$ and $Y$ can be calculated as:
\begin{equation}
\label{eqn:pred}
    {\hat{z}} = f_{\mathrm{pred}}(\mathbf{e}^{\mathrm{re}} (X), \mathbf{e}^{\mathrm{re}}(Y)),
\end{equation}
where $f_{\mathrm{pred}}$ can be any of the underlying models such as Sentence-BERT~\cite{reimers2019sentence}, Lawformer~\cite{xiao2021lawformer}, Bert-PLI~\cite{shao2020bert}, IOT-Match~\cite{yu2022explainable} etc. 

\subsection{Model training}
\label{sec: training}
Law-Match has parameters to learn, including those in treatment reconstruction module for $X$ (i.e., parameters in $f^s_{\mathrm{proj}}$, $f^s_{\mathrm{attn}}$, and $f^s_{\mathrm{weight}}$) and those in treatment reconstruction module for $Y$. We denote these parameters as $\Theta_1$. The underlying matching model $f_{\mathrm{pred}}$ also has another set of learn-able parameters, denoted as $\Theta_2$. 
Law-Match designs an alternative optimization procedure for learning these parameters based on the labelled training data $\mathcal{D}$. 
Each optimization iteration consists of two stages: the IV regression stage for updating $\Theta_1$ and the matching stage for updating $\Theta_2$. 

At each batch of the IV regression stage, after sampling $n$ training pairs and the cited law articles, IV regression is employed to output the law-related representations $\widehat{\mathbf{e}}(X_i)$ and $\widehat{\mathbf{e}}(Y_i)$ for all $i =1, \cdots, n$. MSE (mean square error) is used to measure the losses during the IV regression stage,
\begin{equation}
\label{eqn:loss1}
    \mathcal{L}_{\mathrm{IV}}=\frac{1}{n}\sum\limits_{i=1}^n\left\{\|\widehat{\mathbf{e}}(X_i)-\mathbf{e}(X_i)\|^2+\|\widehat{\mathbf{e}}(Y_i)-\mathbf{e}(Y_i)\|^2\right\}.
\end{equation}
Gradients are then calculated to update the parameters in $\Theta_1$. 

Moving to the matching stage and at each batch, after sampling $n$ training pairs and the cited law articles, the predicted matching scores $\hat{z}_i$'s are calculated according to current parameter values. Cross entropy is employed to measure the matching loss:
\begin{equation}
\label{eqn:loss2}
    \mathcal{L}_{\mathrm{match}} = \frac{1}{n}\sum\limits_{i=1}^n \mathrm{CE}(\widehat{z}_i, {z}_i),
\end{equation}
where CE$(\widehat{z}_i, {z}_i)$ denotes the cross-entropy between the prediction $\widehat{z}_i$ and the ground-truth label $z_i$. Gradients are calculated to update the underlying matching model's parameters $\Theta_2$.

\section{Discussion}
In real-world applications, Law-Match may face several issues. The following subsections discuss how to address them.   
\subsection{Missing of the query law articles}\label{subsec:MissLawArticle}
    


In some real tasks such as legal retrieval, the law articles may not be provided in the source case $X$. One reason is that the source cases (queries) are not judged yet. 
To address the issue, we adapt the causal discovery method proposed in~\cite{liueverything} to predict the missing law articles in the legal case\footnote{Note that causal discovery in~\cite{liueverything} is originally designed for the task of similar charge disambiguation. Two modifications are made to adapt for predicting law articles: replacing the charge names with law articles and changing the operation level from the terms to sentences.}.

The causal discovery method consists of two steps: (1) constructing a bipartite graph $\mathcal{G}$ for describing the relation between legal cases and law articles based on large-scale legal cases with cited law articles; and (2) inferring the law articles for the given legal cases based on $\mathcal{G}$.

The first step aims to create a bipartite graph $\mathcal{G}=(U, V, E)$ where $U$ is the set of $C$ sentence clusters, $V$ is the set of $M$ law article IDs, and $E$ is the set of edges from $V$ to $U$. The graph is created based on 100,000 legal cases with law articles, crawled from \url{https://www.faxin.cn}. Each case contains sentences describing the facts and a set of cited law articles IDs. Following the practices in the Section 3 of~\cite{liueverything}, $\mathcal{G}$ can be created by selecting key sentences from each legal case, clustering the key sentences, filtering unimportant sentences associated with the law articles, and finally creating links ($E$) that link the law article IDs ($U$) to clusters IDs ($V$). Please refer to~\cite{liueverything} for the detailed procedure.

In the second step, given the legal case $X$ and graph $\mathcal{G}$, the law articles can be  predicted as follows:

     (1) Split $X$ into sentences $x_1, \cdots, x_K$;
     
     (2) Assigning sentences $x_i$ ($i =1 ,\cdots, K$) to the clusters in $U$, using the embeddings generated by a pre-trained LMs and Euclidean distance. Note one sentence may be assigned to multiple clusters. 
     
     (3) For each cluster, selecting at most $K'$ assigned sentences, according to the summed distances between sentences and $U$.
     
     (4) For each $x_i (i=1,\cdots,K)$, collecting the set of associated law article IDs $\mathcal{A}_i$ (i.e., moving one step, starting from the assigned clusters (nodes in $U$) and following the edges in $E$);
     
     (5) Return $\bigcup_{i=1}^K \mathcal{A}_i$.






\subsection{\mbox{Underlying models that use paragraph inputs}}\label{subsec:ParagraphInput}
Some legal case matching models, such as Bert-PLI~\cite{shao2020bert} require paragraph embeddings, not the document embeddings, as inputs. For these models, Law-Match considers each paragraph of the legal case as a (pseudo) legal case. Therefore, each paragraph can be processed individually, achieving a single reconstructed vector. Let us use Law-Match, which uses BERT-PLI as the underlying model, as an example. Suppose that in $(X, Y)$, $X$ contains $m$ paragraphs and $Y$ contains $n$ paragraphs. Law-Match will apply its treatment reconstruction $m\times n$ times. Each time, it takes a different paragraph pair (constructed based on $X$ and $Y$) as the input, generating a pair of reconstructed paragraph embeddings. After that, these $m\times n$ embedding pairs are fed to BERT-PLI for conducting the final matching prediction.   

Note that the paragraphs in a legal case have no law articles associated because the law articles are usually cited at the end of legal cases. Law-Match predicts the law articles for each paragraph using the causal discovery described in Section~\ref{subsec:MissLawArticle}. Specifically, (1) if the original legal case also cites no law articles, for each paragraph in the case, we directly use the procedure in Section~\ref{subsec:MissLawArticle}, and return the predicted law articles; (2) if the original legal case cites law articles, we first use the cited law articles to prune the graph $\mathcal{G}$, i.e., removing those nodes $V$ (law articles IDs) that are not cited by the original legal case. Then, we use the procedure in Section~\ref{subsec:MissLawArticle} with the reduced graph to predict law articles for each paragraph. 




\subsection{Feasibility of using law articles as IVs}


This section discusses whether law articles are valid IVs for legal case matching. A valid IV need to satisfy three assumptions: \textit{Relevance}, \textit{Exclusion Restriction} and \textit{Instrumental Unconfoundedness}~\cite{peters2017elements}. 

As for \textit{Relevance}, it means that the IVs (law articles) need to be relevant to the treatments (legal cases).
In order to verify the relevance, we use distance correlation (dCor)~\cite{szekely2014partial}, which measures linear and nonlinear associations between two random variables, to measure the relevance between treatments and IVs. dCor$\in[0,1]$ where a larger dCor means more relevant. For each legal case in ELAM (details  Section~\ref{sec:dataset}), we calculate the dCors value between the cited law articles and the fact descriptions of the legal case. The averaged dCor$=0.7327$. As for comparison, we replace the cited law articles with law articles randomly selected from a law book. The averaged dCor decreases to dCor$=0.2673$. The result indicates that the cited law articles satisfy the relevance assumption. 

As for \textit{Exclusion Restriction}, the IVs causal effect on the outcome is fully
mediated by the treatment. Law-Match estimates the similarities between the source and target cases. The law articles are associated to concrete legal cases rather than the matching labels. This means the law articles can only effect the matching labels through concrete legal cases. Therefore, Law-Match meets the \textit{Exclusion Restriction} assumption.

As for \textit{Instrumental Unconfoundedness}, the IVs need to be uncorrelated with the confounders. In Law-Match, the confounders could be some missing variables (e.g., the focus of disputes) unrelated to any law articles. 
The law articles only affect the key elements in the legal case. Therefore, the law articles are independent of the confounder, satisfying the exogeneity assumption. 


\section{Experiments}
In this section, we empirically verify the efficiency of Law-Match. The source code and all experiments have been shared at ~\url{https://github.com/Jeryi-Sun/Law-Match-SIGIR-23}.






\subsection{Experimental settings}
\subsubsection{Datasets}
\label{sec:dataset}
The experiments were conducted based on three publicly available datasets: ELAM~\cite{yu2022explainable}, eCAIL~\cite{yu2022explainable}, and LeCaRD~\cite{ma2021lecard}. 

\textbf{ELAM} is an explainable legal case matching dataset. It contains 1250 source legal cases, each associated with four target cases. Each legal case pair is manually assigned a matching label which is either match (2), partially match (1), or mismatch (0). Explainable labels such as rationales, their alignments, and free-form explanations are also provided in the dataset. In the experiments, we use the legal case contents and the matching labels for training and evaluating the matching models.  

\textbf{eCAIL} is an extension of CAIL (Challenge of AI in Law) 2021 dataset\footnote{Fact Prediction Track data: \url{http://cail.cipsc.org.cn/}}. In CAIL data, each legal case is associated with tags about private lending. Following the  practices in~\cite{yu2022explainable}, we constructed 1875 source cases, each associated with four target cases. Each legal case pair is assigned a matching label according to the number of overlapping tags (match if overlapping $>10$, mismatch if $<1$, and partially match otherwise).

\textbf{LeCaRD} is a legal case retrieval dataset which contains 107 source (query) cases and 43,000 target cases. All criminal cases were published by the Supreme People's Court of China. For each query, 30 target cases are manually annotated, each assigned a 4-level relevance (matching) label.


As for the law articles, we count each dataset's occurrences of different law categories. Specifically, for LeCaRD and ELAM, we use the PRC Criminal Law; for eCAIL, we use the PRC Contract Law and Civil Procedure Law. The contents of law articles are downloaded from \url{https://flk.npc.gov.cn/}.


    


\subsubsection{Baselines and evaluation metrics}
\label{sec:baseline models}
The proposed Law-Match is a model-agnostic framework, which is applied to the following underlying models:

\textbf{Sentence-BERT}~\cite{reimers2019sentence} is a text matching model. It uses BERT~\cite{devlin2018bert} to encode two sentences separately. Then it concatenates the two embeddings together and uses an MLP to conduct matching.

\textbf{Lawformer}~\cite{xiao2021lawformer} is a Longformer-based pre-trained language model training millions of Chinese legal cases to represent long legal documents better. In the experiment, we send the texts of two cases together to Lawformer and use the mean pooling of Lawformer’s output to conduct matching.

\textbf{BERT-PLI}~\cite{shao2020bert} uses BERT to capture the semantic relationships at the paragraph level. Then it uses RNN and Attention model to infer the relevance between the two cases. Finally, it uses an MLP to calculate the aggregated embeddings similarity score.

\textbf{IOT-Match}~\cite{yu2022explainable} is a three-stage model designed for explainable legal case matching. It extracts rationales in the first stage, generates natural language-based explanations in the second stage, and conducts explainable legal case matching in the third stage. We kept the first two stages identical to~\cite{yu2022explainable} and conducted Law-Match at the third stage. Note that IOT-Match needs explainable features which are unavailable in LeCaRD, we only compared with IOT-Match on ELAM and eCAIL.

We also compare Law-Match with two baselines that can also add the law articles' knowledge to legal cases. The first is an intuitive baseline that simply appends the contents of cited law articles to the original cases, forming new extended legal cases. Existing matching models of Sentence-BERT, Lawformer, BERT-PLI and IOT-Match can be applied to the extended legal cases, denoted as \textbf{Cat-Law (Sentence-BERT)}, \textbf{Cat-Law (Lawformer)}, \textbf{Cat-Law (BERT-PLI)}, and \textbf{Cat-Law (IOT-Match)}, respectively. 
The second is from~\cite{feng-etal-2022-legal} which employs an attention mechanism to incorporate article semantics into the legal judgement prediction models called EPM. Existing matching models of Sentence-BERT, Lawformer, BERT-PLI and IOT-Match can be applied to EPM, denoted as \textbf{EPM (Sentence-BERT)}, \textbf{EPM (Lawformer)}, \textbf{EPM (BERT-PLI)}, and \textbf{EPM (IOT-Match )}, respectively.

The proposed Law-Match is model-agnostic. In the experiments, we applied Law-Match to the baselines of  Sentence-BERT, Lawformer,  BERT-PLI, and IOT-Match achieving four versions, referred to as \textbf{Law-Match (Sentence-BERT)}, \textbf{Law-Match (Lawformer)},  \textbf{Law-Match (BERT-PLI)}, and \textbf{Law-Match (IOT-Match )} respectively.

As for evaluation metrics, we use Accuracy, Macro-Precision, Macro-Recall, and Macro-F1 to measure the matching accuracy. 
\subsubsection{Implementation details}
Law-Match's hyperparameters are tuned using grid search on the validation set with Adam~\cite{kingma2015method}. 
The batch size is tuned among $\{2, 4, 8\}$. The learning rate $\eta_1$ and $\eta_2$ are tuned among $\{3e-6, 3e-5, 3e-4\}$. For the source cases that do not cite law articles, the number of  automatically  discovered law articles $K'$ is tuned between [3, 15] with step 2. For baselines, we set the parameters as the optimal values in the original paper.
 
We use Legal-Bert\footnote{\url{https://github.com/thunlp/OpenCLaP}} to encode legal cases (and the corresponding law articles if needed) for Sentence-BERT~\cite{reimers2019sentence}, Bert-PLI~\cite{shao2020bert}, and IOT-Match~\cite{yu2022explainable}. 
The legal cases from eCAIL are generally longer than BERT's maximum input length. For Sentence-BERT~\cite{reimers2019sentence}, we use TextRank~\cite{mihalcea2004textrank} to process the legal cases and generate a summary with a 512-words for each case. For Bert-PLI~\cite{shao2020bert} and  Lawformer~\cite{xiao2021lawformer}, the original text is used. 

\subsection{Experimental results and analysis}

\subsubsection{\textbf{Comparison against underlying models and baselines}}
\begin{table*}[t]
\setlength{\abovecaptionskip}{0.3cm}
 \caption {Performance comparisons between Law-Match and the baselines. The boldface represents the best performance. In each block, we present the Law-Match in the last line. `$\dagger$' indicates the improvements over all of the baselines are  statistically significant (t-tests, $p\textrm{-value}< 0.05$). Results of IOT-Match on LeCaRD are not available, denoted as `---'. 
}\label{main results} 
\centering
\resizebox{0.95\linewidth}{!}{
\begin{tabular}{lcccccccccccc}
\toprule

\multirow{2}{*}{\textbf{Models}} & \multicolumn{4}{c}{\textbf{ELAM}} & \multicolumn{4}{c}{\textbf{LeCaRD}} & \multicolumn{4}{c}{\textbf{eCAIL}}                                                                      \\
\cmidrule(lr){2-5}\cmidrule(lr){6-9}\cmidrule(lr){10-13}
  & \textbf{Acc. (\%)}& \textbf{P. (\%)}& \textbf{R. (\%)} &  \textbf{F1 (\%)}    &  \textbf{Acc. (\%)} & \textbf{P. (\%)}& \textbf{R. (\%)} & \textbf{F1 (\%)}   & \textbf{Acc. (\%)}  &  \textbf{P. (\%)}& \textbf{R. (\%)} & \textbf{F1 (\%)}  \\
\midrule
 Sentence-BERT &  68.83 &69.83 &66.88 &67.20 &59.44 & 59.54
&57.89 & 58.70 &  71.33 &70.83 &71.21 &70.98                \\
Cat-Law(Sentence-BERT) &  71.54 &70.54  &69.73 & 69.94 & 61.60 & 62.54	& 59.76&	60.73& 78.80 & 78.36&78.70&78.53     \\
EPM(Sentence-BERT) &  71.14&69.85&69.51&69.65&60.37&61.56&58.12&59.25&77.06&76.65&76.95&76.58\\
Law-Match(Sentence-BERT) &  $\textbf{73.15}^\dagger$& $\textbf{71.23}^\dagger$ & $\textbf{71.05}^\dagger$ & $\textbf{71.14}^\dagger$ & $\textbf{62.54}^\dagger$	& $\textbf{63.37}^\dagger$& $\textbf{61.04}^\dagger$	& $\textbf{61.84}^\dagger$ & $\textbf{80.00}^\dagger$ & $\textbf{79.78}^\dagger$ & $\textbf{79.92}^\dagger$& $\textbf{79.84}^\dagger$
  \\
\midrule
Lawformer &69.91&72.26 &68.34&69.18&59.13&58.79&58.56&58.47&70.67 &70.20 &70.55 &69.91\\
Cat-Law(Lawformer) &  69.94&68.05&68.40&68.22&59.13&59.27&58.54&58.59&75.19&75.51&75.18&75.20 \\
EPM(Lawformer) & 71.14	&72.75&70.58&70.31&59.37&60.02&59.14&59.43&74.00&73.85&74.20&74.00\\
 Law-Match(Lawformer) &
  \textbf{73.20}$^\dagger$&\textbf{74.41}$^\dagger$&\textbf{73.12}$^\dagger$&\textbf{73.52}$^\dagger$&\textbf{60.06}$^\dagger$&\textbf{60.80}$^\dagger$&\textbf{59.54}$^\dagger$&\textbf{59.62}$^\dagger$&
 \textbf{76.67}$^\dagger$&\textbf{76.25}$^\dagger$&\textbf{76.56}$^\dagger$&\textbf{76.40}$^\dagger$ \\           
 \midrule
BERT-PLI & 71.21 &71.22 &71.23 &70.88&61.60&60.88	&60.41&60.48&70.66 &70.05 &70.54 &70.18
 \\
Cat-Law(BERT-PLI) &  72.89&71.32&70.49&70.63&63.46&64.15&62.47&63.16&73.20&72.51	&73.08&72.28
 \\
 EPM(BERT-PLI) &71.34&69.32&69.11&68.99&63.77&65.26&62.45&63.49&73.33&73.12&73.23&73.18 \\
 Law-Match(BERT-PLI) &  \textbf{74.95}$^\dagger$&\textbf{72.96}$^\dagger$&\textbf{71.75}$^\dagger$&\textbf{72.35}$^\dagger$&\textbf{65.63}$^\dagger$&	\textbf{66.07}$^\dagger$&\textbf{63.75}$^\dagger$&\textbf{64.41}$^\dagger$
 &\textbf{74.13}$^\dagger$&\textbf{73.51}$^\dagger$&\textbf{74.02}$^\dagger$&\textbf{73.68}$^\dagger$
  \\
\midrule
IOT-Match & 73.87&73.02&72.41&72.55&---&---&---&---&82.01&82.10&81.92&81.90             \\
Cat-Law(IOT-Match) & 74.55&73.22&72.63&72.89&---&---&---&---&83.86&84.59&83.72&83.95 \\
 EPM(IOT-Match) & 74.69&73.39&73.02&73.17&---&---&---&---&82.53&82.21&82.40&82.29 \\
 Law-Match(IOT-Match) &
  \textbf{76.75}$^\dagger$&\textbf{75.51}$^\dagger$&\textbf{75.78}$^\dagger$&\textbf{75.59}$^\dagger$&---&---&---&---&
  \textbf{84.60}$^\dagger$&\textbf{84.44}$^\dagger$&\textbf{84.53}$^\dagger$&\textbf{84.45}$^\dagger$ \\         
\bottomrule
\end{tabular}
}
\end{table*}

From the results reported in Table~\ref{main results}, we found that Law-Match (Sentence-BERT), Law-Match (Lawformer), and Law-Match (BERT-PLI) outperformed the corresponding underlying models (i.e., Sentence-BERT, Lawformer, and BERT-PLI) on all of the three datasets (ELAM, eCAIL, and LeCaRD) and Law-Match (IOT-Match) outperformed IOT-Match on ELAM and eCAIL, with statistical significance (t-tests, $p$-value <0.05). The results verified the efficiency of the model-agnostic Law-Match framework in improving the underlying matching models.

Meanwhile, we find that the four versions of Cat-Law/EPM, i.e., Cat-Law/EPM (Sentence-BERT), Cat-Law/EPM (Lawformer), Cat-Law/EPM (BERT-PLI), and Cat-Law/EPM (IOT-Match) also outperform most of the underlying models, indicating that the knowledge from the law articles helps improve legal case matching. 
Finally, Law-Match (Sentence-BERT), Law-Match (Lawformer), Law-Match (BERT-PLI), and Law-Match(IOT-Match) outperform the corresponding baselines of Cat-Law/EPM (Sentence-BERT), Cat-Law/EPM (Lawformer), Cat-Law/EPM (BERT-PLI), and Cat-Law/EPM (IOT-Match), verify that considering law articles as IVs to decompose treatments is a better way of using law articles in legal cases matching. 


%
\begin{table*}[t]
\setlength{\abovecaptionskip}{0.3cm}
\caption {Ablation study of Law-Match on ELAM.
}\label{tab:abs_study IV structure} 
\centering
\resizebox{0.95\linewidth}{!}{
\begin{tabular}{lcccccccccccccccc}
\toprule
\multirow{2}{*}{\textbf{Algorithm}} & \multicolumn{4}{c}{\textbf{Sentence-BERT}} & \multicolumn{4}{c}{\textbf{Lawformer}} & \multicolumn{4}{c}{\textbf{Bert-PLI}} & \multicolumn{4}{c}{\textbf{IOT-Match}}                                                                   \\
\cmidrule(lr){2-5}\cmidrule(lr){6-9}\cmidrule(lr){10-13}\cmidrule(lr){14-17}
  & \textbf{Acc. (\%)}& \textbf{P. (\%)}& \textbf{R. (\%)} &  \textbf{F1 (\%)}    &  \textbf{Acc. (\%)} & \textbf{P. (\%)}& \textbf{R. (\%)} & \textbf{F1 (\%)}   & \textbf{Acc. (\%)}  &  \textbf{P. (\%)}& \textbf{R. (\%)} & \textbf{F1 (\%)} &\textbf{Acc. (\%)}  &  \textbf{P. (\%)}& \textbf{R. (\%)} & \textbf{F1 (\%)} \\
\midrule
Law-Match (fitted
only) & 66.53&66.47&64.69&65.15
 &71.74&69.90&69.33&69.35 &  70.74&68.71&68.35&68.52&70.74&69.89&69.31&69.45
\\
 Law-Match (residual only) &  69.13	&68.01	& 68.19&68.10&71.14&69.37&68.68&68.70&69.53&67.05&66.19&66.62& 73.34&71.83&72.33&71.96\\
  Law-Match (Concat parts) & 71.74&70.55&70.65&70.54&
 71.03&70.87&70.56&70.34 & 73.55&71.52&71.68&71.52&74.95&74.11&74.31&74.20\\
 Law-Match (Separate IV) & 72.55&71.05&70.73&70.82&
 71.14&71.00&	69.78&70.12
& 71.74	&69.15&68.72&68.93&74.54&73.14&69.63&69.30
 \\
 \midrule
 Law-Match  &  \textbf{73.15}& \textbf{71.23} & \textbf{71.05} &\textbf{71.14} &\textbf{73.20}&\textbf{74.41}&\textbf{73.12}&\textbf{73.52}&\textbf{74.95}&\textbf{72.96}&\textbf{71.75}&\textbf{72.35}& \textbf{76.75}&\textbf{75.51}&\textbf{75.78}&\textbf{75.59} \\
\bottomrule
\end{tabular}
}
\end{table*}
\subsubsection{\textbf{How the law articles improve legal case matching? }}

\begin{figure}[t]
    \subfigure[average distance: 13.20]{
    \includegraphics[width=0.44\linewidth]{ 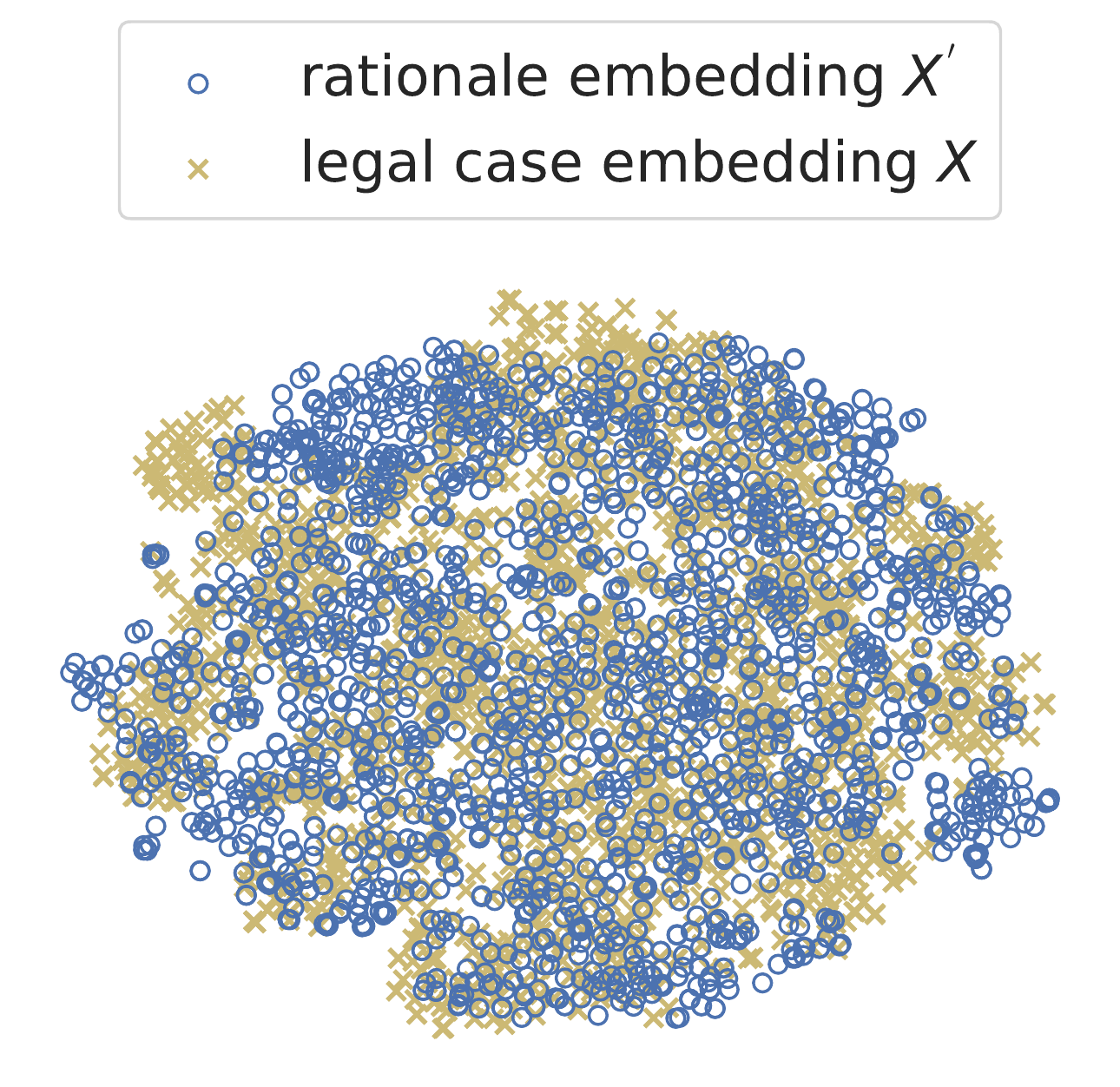}
    \label{fig:base_embedding}
    }
    \subfigure[average distance: 0.28]{
    \includegraphics[width=0.5\linewidth]{ 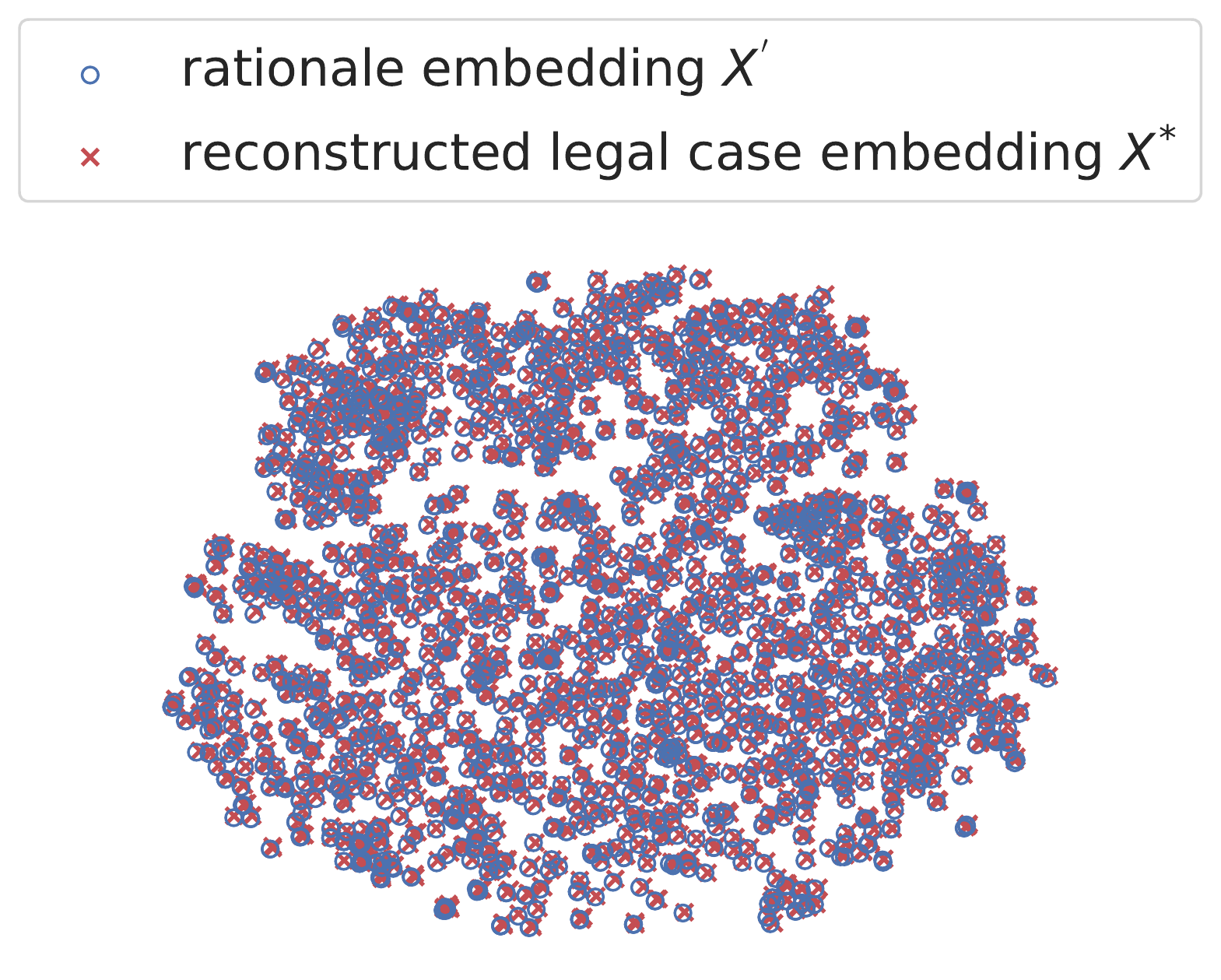}
    \label{fig:IV_embedding}
    }
    \caption{
    (a): Embeddings  of legal case generated by Legal-Bert v.  Embeddings  of rationale generated by Legal-Bert. (b): Embeddings  of legal case generated by Legal-Bert v.  Reconstructed Embeddings of legal case generated by Law-Match.}
    \label{fig:rationale_similarity}
\end{figure}

We first show that Law-Match has the ability of \textit{Identifying Rationales.} That is, the law articles guide Law-Match to reconstruct the legal case embeddings that focus more on the rationales, which have been verified to be beneficial to accurate matching~\cite{yu2022explainable}. Specifically, we note that each legal case in ELAM also contains human-annotated rationales (key sentences). 
Therefore, for each legal case, we generate the legal case embedding $X$ by Legal-Bert, the rationale embedding $X'$ which contains only the human-annotated rationales by Legal-Bert, and the reconstructed legal cases embedding $X^{*}$ by Law-Match.

First, we use TSNE~\cite{van2008visualizing} to illustrate all of the 2500 legal case embeddings in~\autoref{fig:rationale_similarity}. ~\autoref{fig:rationale_similarity}(a) shows the distributions of the legal case embeddings $X$ (yellow crosses) and the rationale embeddings $X'$ (blue circles). ~\autoref{fig:rationale_similarity}(b) shows the distributions of the reconstructed legal case embeddings ${X}^{*}$ (red crosses) and the rationale embeddings $X'$ (blue circles). It is easy to observe that the blue circles and yellow crosses in \autoref{fig:rationale_similarity}(a) are distributed more differently than the blue circles and red crosses in \autoref{fig:rationale_similarity}(b). That is, more circles and crosses are not overlapped in \autoref{fig:rationale_similarity}(a). 
Moreover, we calculate the averaged Euclidean distances over all of the pairs $(X, X')$ in \autoref{fig:rationale_similarity}(a) based on embedding vectors generated by Legal-Bert. The average distance is 13.20. After applying Law-Match on the legal cases and based on the reconstructed embeddings, the average Euclidean distances over all of the pairs $(X^{*}, X')$ in \autoref{fig:rationale_similarity}(b) becomes 0.28. 
The results verify that by using law articles as IVs, Law-Match reconstruct the case embeddings so that they are closer to the corresponding rationale embeddings.

\begin{figure}[t]
    \subfigure[Law-Match(Sentence-BERT)]{
    \includegraphics[width=0.45\linewidth]{ 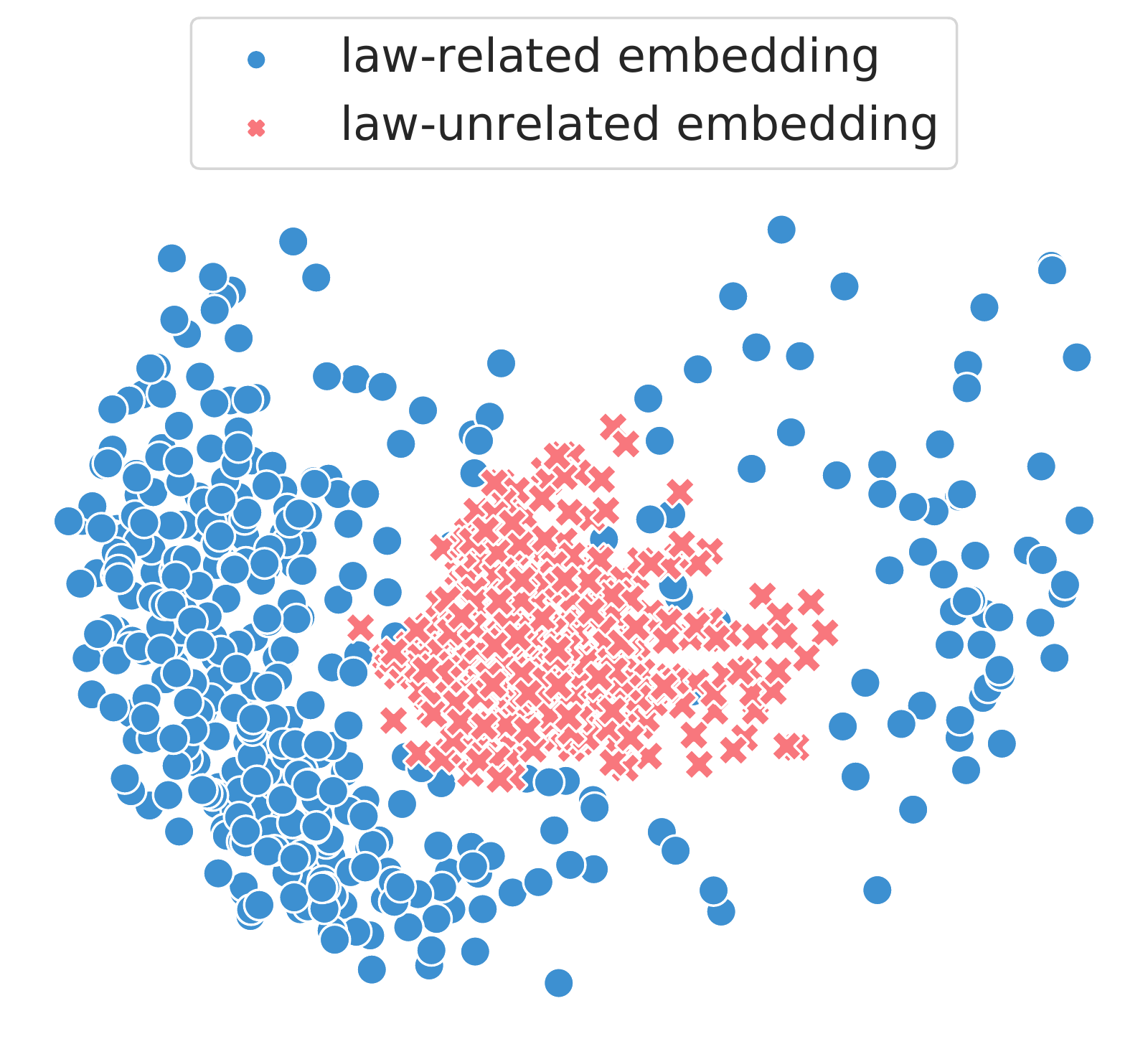}
    \label{fig:Sentence-BERT embedding}
    }
    \subfigure[Law-Match(Lawformer)]{
    \includegraphics[width=0.45\linewidth]{ 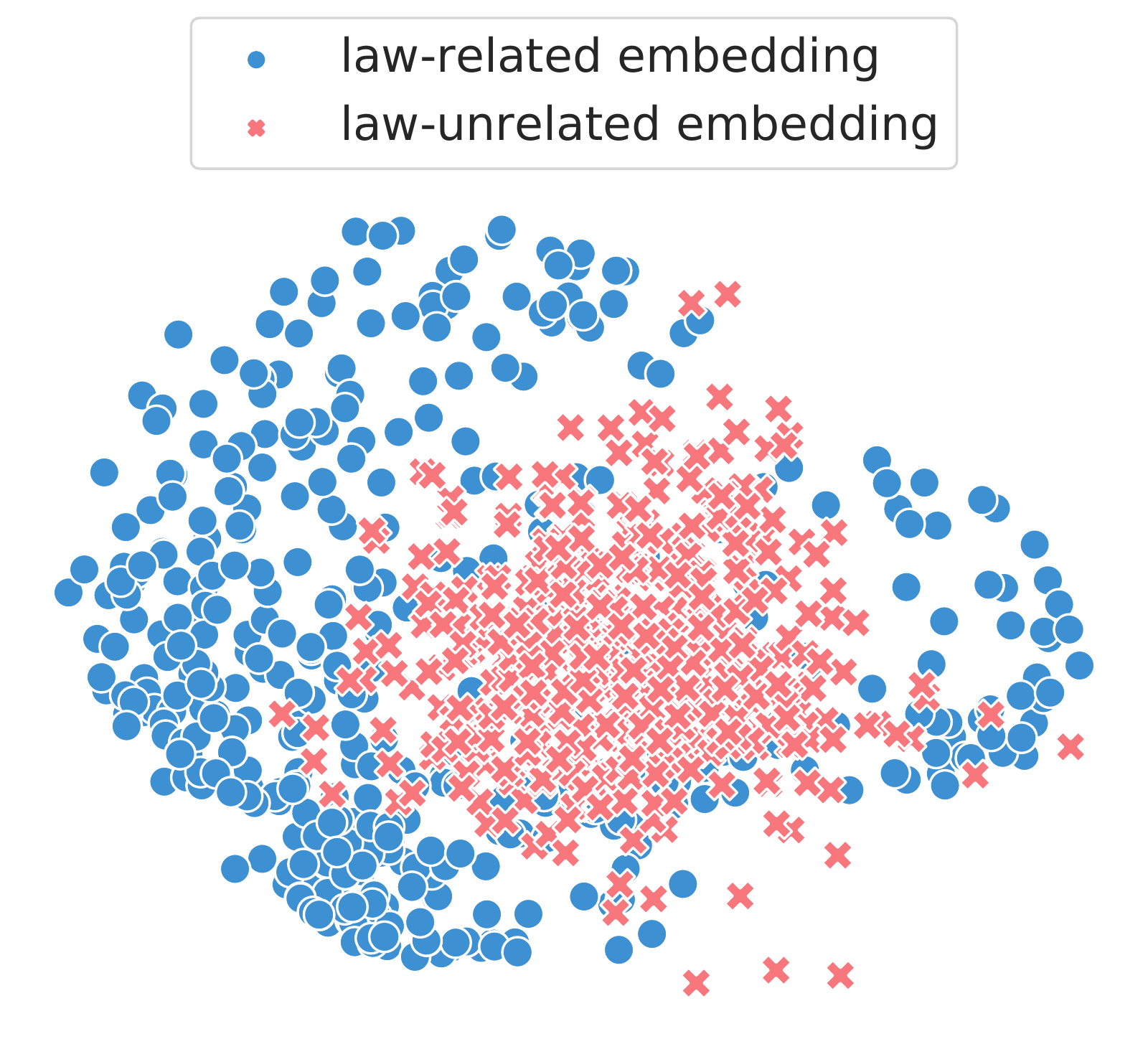}
    \label{fig:Lawformer embedding}
    }
    \caption{
    Visualization of the learned causal and non-causal embeddings of (a) Law-Match(Sentence-BERT) and (b) Law-Match(Lawformer). Causal parts are represented by dots and non-causal parts are represented by crosses. Using law articles as IVs, causal and non-causal parts are disentangled clearly by Law-Match.}
    \label{fig:embedding of causal and non-causal}
\end{figure}

We further show that Law-Match has the ability of \textit{disentangling the law-related and law-unrelated parts} of treatment vectors in Section~\ref{sec: main model}. 
Specifically, we visualize the decomposed law-related and law-unrelated embeddings learned by Law-Match using TSNE. Figure~\ref{fig:embedding of causal and non-causal}(a) and (b) show the results w.r.t. Law-Match(Sentence-BERT) and Law-Match(Lawformer), respectively. The law-related embeddings are shown as blue dots, and the law-unrelated embeddings are shown as red crosses. In Figure~\ref{fig:embedding of causal and non-causal}, we observe that Law-Match separates the two sets of embeddings. Only a tiny fraction of the vectors are overlapped. From the above analysis, we conclude that Law-Match effectively disentangles the law-related and law-unrelated parts of treatment vectors, which are utilized differently and enhanced legal case matching.

\subsubsection{\textbf{Ablation study}}
Law-Match combines the fitted and residual parts as the reconstructed legal case representation. We create several Law-Match variations by removing the two parts or changing the combination methods. They are (a) Law-Match (fitted only): only use the fitted part as the reconstructed representation; (b) Law-Match (residual only): only use the residual part as the reconstructed representation. Also, note that in Law-Match, the embeddings of $L_X$ and $L_Y$ are used as IVs to regress both $X$ and $Y$. We created another variation: (c) Law-Match (Separate IV): $L_X$ is only used as the IV for $X$, and $L_Y$ is only used as the IV for $Y$; (d) Law-Match (Concat parts): simply concatenate the fitted part and the residual part as the reconstructed representation. 

 Table~\ref{tab:abs_study IV structure} reports the performance of these variations with different underlying models on the ELAM dataset. The results indicate that (1) removing either the fitted or residual part will decrease the matching performance; (2) combining $L_X$ and $L_Y$'s embeddings as IVs can further enhance the matching accuracy; (3) simply concatenating the fitted part and the residual part will decrease the matching performance. 
\begin{figure}[t]
    \centering
    \includegraphics[width=0.98\linewidth]{ 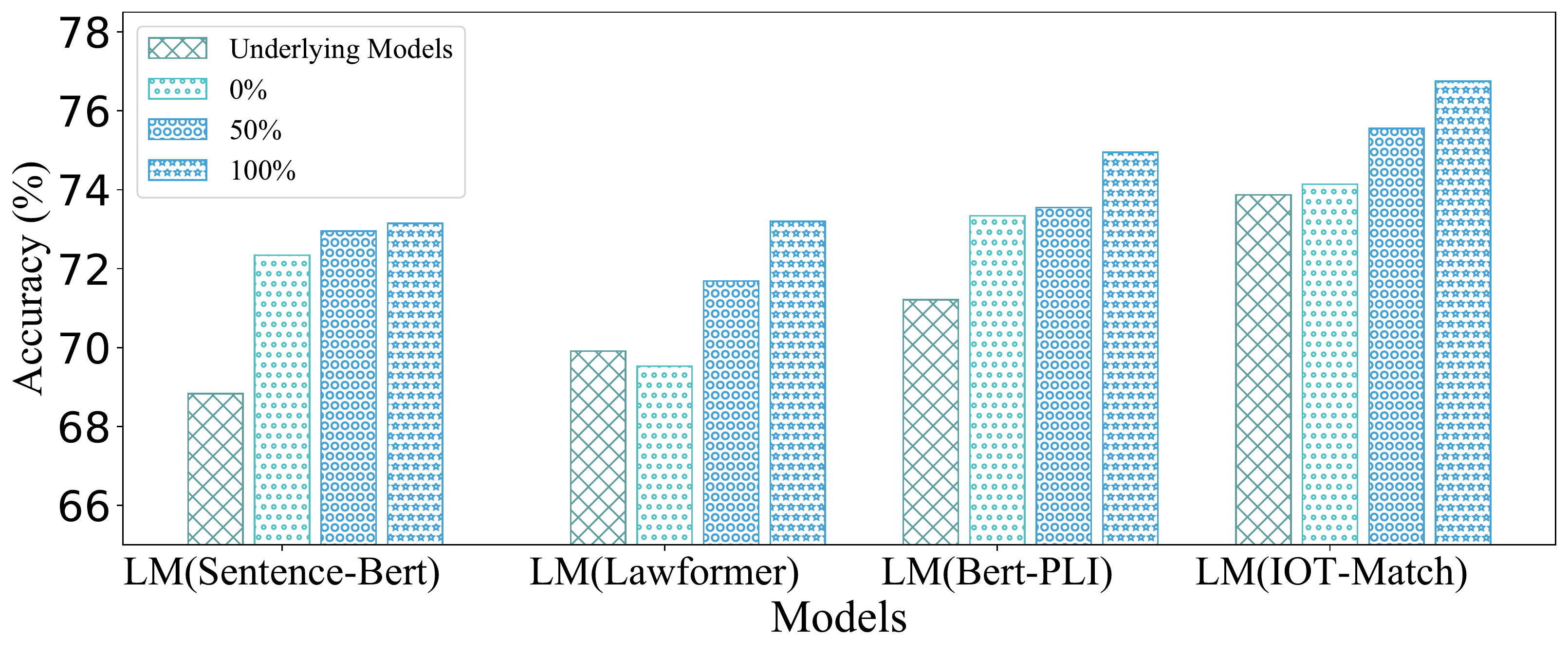}
    \caption{Performance of Law-Match (abbr. LM) w.r.t. different ratio of (oracle) cited law articles on ELAM.}
    \label{fig:laws validity as IVs}
\end{figure}

\subsubsection{\textbf{Robustness of law articles as IVs.}}\label{sec:emp:robust}
Based on the ELAM dataset, we test the performances of Law-Match when a few oracle law articles (those cited in the legal cases) are replaced with those randomly selected from a law book. That is, we try to inject noise into the IVs. Figure~\ref{fig:laws validity as IVs} illustrates the matching accuracy of Law-Match w.r.t. 0\%, 50\%, and 100\% of the oracle law articles are kept (others are replaced with random law articles) and the underlying modes without Law-Match. The results indicate that: (1) Law-Match is robust. It improved the underlying matching models even the law articles are randomly selected; 
(2) high-quality law articles can further enhance the matching accuracy.  

\subsubsection{\textbf{Effects of the causally discovered law articles}}
\begin{figure}[t]
    \centering
    \includegraphics[width=0.98\linewidth]{ 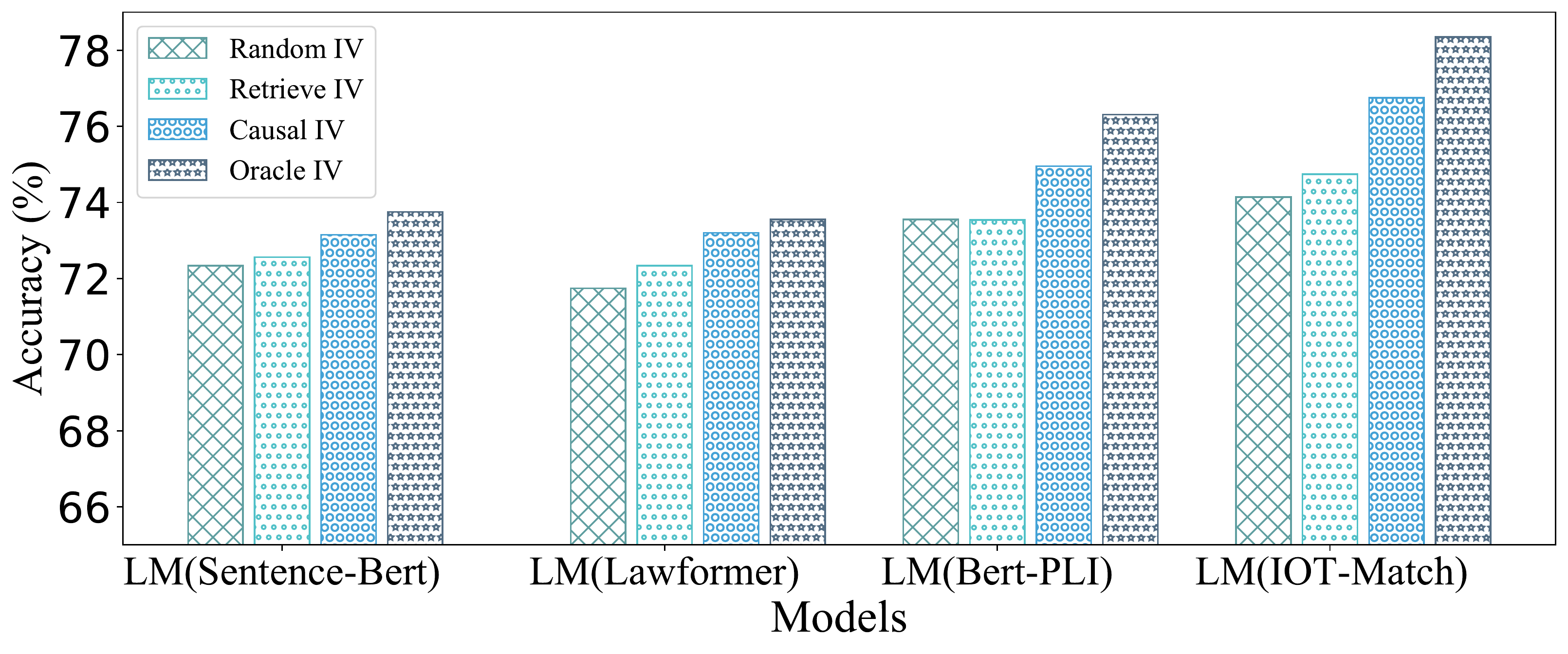}
    \caption{Performance of Law-Match (abbr. LM) w.r.t. law articles discovered by different methods on ELAM. }
    \label{fig:causal discovery effective}
\end{figure}
We test the efficiency of automatically discovered law articles (shown in Section~\ref{subsec:MissLawArticle}) in Law-Match. Specifically, we note that the original legal cases in ELAM contain oracle-cited law articles. We compare the matching accuracy of Law-Match variations where the law articles are collected differently: (1) randomly selecting 5 law articles\footnote{On average each ELAM legal case cites about 5 law articles.}, denoted as Law-Match (Random IV); (2) using BM25~\cite{robertson2009probabilistic} to retrieve top-5 law articles from the law book, where legal cases and law articles are respectively considered as queries and documents, denoted as Law-Match (Retrieve IV); (3) using the causal discovery method presented in \autoref{subsec:MissLawArticle}, denoted as Law-Match (Causal IV); (4) For showing the upper bound performance, we also test Law-Match with oracle cited law articles in the legal cases, denoted as Law-Match (Oracle IV). 

\autoref{fig:causal discovery effective} shows that Law-Match (Causal IV) perform better than Law-Match (Retrieve IV) and Law-Match (Retrieve IV) perform better than Law-Match (Random IV). Law-Match (Oracle IV) perform the best. 
The results verify the efficiency of the causal discovery module presented in \autoref{subsec:MissLawArticle}, especially when no law articles are cited in the legal cases.

\subsubsection{\textbf{Efficiency of Law-Match.}}\label{sec:emp:time analysis}

\begin{table}[h]
    \centering
    \setlength{\abovecaptionskip}{0.3cm}
    \caption{Average online inference time per case pair, for four base models w/ or w/o Law-Math.
    }
    \resizebox{0.95\linewidth}{!}{
    \begin{tabular}{l|c|c|c|c}
        \toprule
        Model&Sentence-Bert&Lawformer&Bert-PLI&IOT-Match\\
        \midrule
         w/o Law-Match & 0.0502 (s) & 0.0318 (s) & 0.1228 (s) & 0.1027 (s)\\
         w/ Law-Match& 0.0543 (s) & 0.0341 (s)& 0.1269 (s)& 0.1059 (s) \\
         \midrule
         RelaCost & 8.17 \% & 7.23 \% & 3.34 \% & 3.12 \% \\
         \bottomrule
    \end{tabular}
    }
    \label{tab:time analysis}
\end{table}

We also analyze the efficiency of Law-Match to show the additional time needed when conducting online matching. Specifically, we record the time required to process each case pair in the online inference stage, with different base models with and without the Law-Match module. From the results reported in Table~\ref{tab:time analysis}, we find that Law-Match needs a short additional time when applied to different base models. Also, we find that relative additional time costs are lower for the larger base models. 
Overall, the delay is acceptable and will not impact the online inference speed much. 


The results are reasonable because Law-Match is designed as an independent representation learning module in the process of legal matching. The most time-consuming part is reconstructing the treatment, which consists of four MLP layers ($f_{\mathrm{weight}}^{s/t}$, $f_\mathrm{proj}^{s/t}$) and two additive attention networks ($f_{\mathrm{attn}}^{s/t}$). These modules require much less time than the base models.

\section{CONCLUSIONS}
In this paper, we propose a model-agnostic causal learning framework that introduces law articles to legal case matching, called Law-Match. Analyses show that the legal case matching results are affected by the mediation effect of the cited law articles and the direct effect of the key circumstances in legal cases. By considering the law articles as IVs and legal cases as treatments, Law-Match uses IV regression to decompose each legal case's embedding into the law-related and law-unrelated parts, which are then combined together for the final matching prediction. Experiments on three public datasets demonstrated the efficiency of Law-Match.

\begin{acks}
This work was funded by the National Key R\&D Program of China (2019YFE0198200), Beijing Outstanding Young Scientist Program NO. BJJWZYJH012019100020098, Fundamental Research Funds for the Central Universities, and the Research Funds of Renmin University of China (23XNH024), Major Innovation \& Planning Interdisciplinary Platform for the ``Double-First Class'' Initiative, and Public Computing Cloud, Renmin University of China. The work was partially done at Beijing Key Laboratory of Big Data Management and Analysis Methods.
\end{acks}

\bibliographystyle{ACM-Reference-Format}
\bibliography{sample-base}


\begin{thebibliography}{49}


\ifx \showCODEN    \undefined \def \showCODEN     #1{\unskip}     \fi
\ifx \showDOI      \undefined \def \showDOI       #1{#1}\fi
\ifx \showISBNx    \undefined \def \showISBNx     #1{\unskip}     \fi
\ifx \showISBNxiii \undefined \def \showISBNxiii  #1{\unskip}     \fi
\ifx \showISSN     \undefined \def \showISSN      #1{\unskip}     \fi
\ifx \showLCCN     \undefined \def \showLCCN      #1{\unskip}     \fi
\ifx \shownote     \undefined \def \shownote      #1{#1}          \fi
\ifx \showarticletitle \undefined \def \showarticletitle #1{#1}   \fi
\ifx \showURL      \undefined \def \showURL       {\relax}        \fi
\providecommand\bibfield[2]{#2}
\providecommand\bibinfo[2]{#2}
\providecommand\natexlab[1]{#1}
\providecommand\showeprint[2][]{arXiv:#2}

\bibitem[Angrist and Imbens(1995)]%
        {angrist1995two}
\bibfield{author}{\bibinfo{person}{Joshua~D Angrist} {and}
  \bibinfo{person}{Guido~W Imbens}.} \bibinfo{year}{1995}\natexlab{}.
\newblock \showarticletitle{Two-stage least squares estimation of average
  causal effects in models with variable treatment intensity}.
\newblock \bibinfo{journal}{\emph{Journal of the American statistical
  Association}} \bibinfo{volume}{90}, \bibinfo{number}{430}
  (\bibinfo{year}{1995}), \bibinfo{pages}{431--442}.
\newblock


\bibitem[Angrist et~al\mbox{.}(1996)]%
        {angrist1996identification}
\bibfield{author}{\bibinfo{person}{Joshua~D Angrist}, \bibinfo{person}{Guido~W
  Imbens}, {and} \bibinfo{person}{Donald~B Rubin}.}
  \bibinfo{year}{1996}\natexlab{}.
\newblock \showarticletitle{Identification of causal effects using instrumental
  variables}.
\newblock \bibinfo{journal}{\emph{Journal of the American statistical
  Association}} \bibinfo{volume}{91}, \bibinfo{number}{434}
  (\bibinfo{year}{1996}), \bibinfo{pages}{444--455}.
\newblock


\bibitem[Bahdanau et~al\mbox{.}(2014)]%
        {bahdanau2014neural}
\bibfield{author}{\bibinfo{person}{Dzmitry Bahdanau},
  \bibinfo{person}{Kyunghyun Cho}, {and} \bibinfo{person}{Yoshua Bengio}.}
  \bibinfo{year}{2014}\natexlab{}.
\newblock \showarticletitle{Neural machine translation by jointly learning to
  align and translate}.
\newblock \bibinfo{journal}{\emph{arXiv preprint arXiv:1409.0473}}
  (\bibinfo{year}{2014}).
\newblock


\bibitem[Bellavia and Valeri(2018)]%
        {bellavia2018decomposition}
\bibfield{author}{\bibinfo{person}{Andrea Bellavia} {and}
  \bibinfo{person}{Linda Valeri}.} \bibinfo{year}{2018}\natexlab{}.
\newblock \showarticletitle{Decomposition of the total effect in the presence
  of multiple mediators and interactions}.
\newblock \bibinfo{journal}{\emph{American journal of epidemiology}}
  \bibinfo{volume}{187}, \bibinfo{number}{6} (\bibinfo{year}{2018}),
  \bibinfo{pages}{1311--1318}.
\newblock


\bibitem[Beltagy et~al\mbox{.}(2020)]%
        {beltagy2020longformer}
\bibfield{author}{\bibinfo{person}{Iz Beltagy}, \bibinfo{person}{Matthew~E
  Peters}, {and} \bibinfo{person}{Arman Cohan}.}
  \bibinfo{year}{2020}\natexlab{}.
\newblock \showarticletitle{Longformer: The long-document transformer}.
\newblock \bibinfo{journal}{\emph{arXiv preprint arXiv:2004.05150}}
  (\bibinfo{year}{2020}).
\newblock


\bibitem[Bench-Capon et~al\mbox{.}(2012)]%
        {2012A}
\bibfield{author}{\bibinfo{person}{T. Bench-Capon}, \bibinfo{person}{Micha
  Araszkiewicz}, \bibinfo{person}{K. Ashley}, \bibinfo{person}{K. Atkinson},
  \bibinfo{person}{F. Bex}, \bibinfo{person}{F. Borges}, \bibinfo{person}{D.
  Bourcier}, \bibinfo{person}{P. Bourgine}, \bibinfo{person}{J.~G. Conrad},
  {and} \bibinfo{person}{E. Francesconi}.} \bibinfo{year}{2012}\natexlab{}.
\newblock \showarticletitle{A history of AI and Law in 50 papers: 25 years of
  the international conference on AI and Law}.
\newblock \bibinfo{journal}{\emph{Artificial Intelligence \& Law}}
  \bibinfo{volume}{20}, \bibinfo{number}{3} (\bibinfo{year}{2012}),
  \bibinfo{pages}{215--319}.
\newblock


\bibitem[Bhattacharya et~al\mbox{.}(2020a)]%
        {bhattacharya2020hier}
\bibfield{author}{\bibinfo{person}{Paheli Bhattacharya},
  \bibinfo{person}{Kripabandhu Ghosh}, \bibinfo{person}{Arindam Pal}, {and}
  \bibinfo{person}{Saptarshi Ghosh}.} \bibinfo{year}{2020}\natexlab{a}.
\newblock \showarticletitle{Hier-spcnet: a legal statute hierarchy-based
  heterogeneous network for computing legal case document similarity}. In
  \bibinfo{booktitle}{\emph{Proceedings of the 43rd International ACM SIGIR
  Conference on Research and Development in Information Retrieval}}.
  \bibinfo{pages}{1657--1660}.
\newblock


\bibitem[Bhattacharya et~al\mbox{.}(2020b)]%
        {bhattacharya2020methods}
\bibfield{author}{\bibinfo{person}{Paheli Bhattacharya},
  \bibinfo{person}{Kripabandhu Ghosh}, \bibinfo{person}{Arindam Pal}, {and}
  \bibinfo{person}{Saptarshi Ghosh}.} \bibinfo{year}{2020}\natexlab{b}.
\newblock \showarticletitle{Methods for computing legal document similarity: A
  comparative study}.
\newblock \bibinfo{journal}{\emph{arXiv preprint arXiv:2004.12307}}
  (\bibinfo{year}{2020}).
\newblock


\bibitem[Caner and Hansen(2004)]%
        {caner2004instrumental}
\bibfield{author}{\bibinfo{person}{Mehmet Caner} {and} \bibinfo{person}{Bruce~E
  Hansen}.} \bibinfo{year}{2004}\natexlab{}.
\newblock \showarticletitle{Instrumental variable estimation of a threshold
  model}.
\newblock \bibinfo{journal}{\emph{Econometric Theory}} \bibinfo{volume}{20},
  \bibinfo{number}{5} (\bibinfo{year}{2004}), \bibinfo{pages}{813--843}.
\newblock


\bibitem[Devlin et~al\mbox{.}(2019)]%
        {devlin2018bert}
\bibfield{author}{\bibinfo{person}{Jacob Devlin}, \bibinfo{person}{Ming{-}Wei
  Chang}, \bibinfo{person}{Kenton Lee}, {and} \bibinfo{person}{Kristina
  Toutanova}.} \bibinfo{year}{2019}\natexlab{}.
\newblock \showarticletitle{{BERT:} Pre-training of Deep Bidirectional
  Transformers for Language Understanding}. In
  \bibinfo{booktitle}{\emph{{NAACL-HLT} {(1)}}}.
  \bibinfo{publisher}{Association for Computational Linguistics},
  \bibinfo{pages}{4171--4186}.
\newblock


\bibitem[Dias et~al\mbox{.}(2022)]%
        {dias2022state}
\bibfield{author}{\bibinfo{person}{Jo{\~a}o Dias}, \bibinfo{person}{Pedro~A
  Santos}, \bibinfo{person}{Nuno Cordeiro}, \bibinfo{person}{Ana Antunes},
  \bibinfo{person}{Bruno Martins}, \bibinfo{person}{Jorge Baptista}, {and}
  \bibinfo{person}{Carlos Gon{\c{c}}alves}.} \bibinfo{year}{2022}\natexlab{}.
\newblock \showarticletitle{State of the Art in Artificial Intelligence applied
  to the Legal Domain}.
\newblock \bibinfo{journal}{\emph{arXiv preprint arXiv:2204.07047}}
  (\bibinfo{year}{2022}).
\newblock


\bibitem[Dippel et~al\mbox{.}(2020)]%
        {dippel2020causal}
\bibfield{author}{\bibinfo{person}{Christian Dippel}, \bibinfo{person}{Andreas
  Ferrara}, {and} \bibinfo{person}{Stephan Heblich}.}
  \bibinfo{year}{2020}\natexlab{}.
\newblock \showarticletitle{Causal mediation analysis in instrumental-variables
  regressions}.
\newblock \bibinfo{journal}{\emph{The Stata Journal}} \bibinfo{volume}{20},
  \bibinfo{number}{3} (\bibinfo{year}{2020}), \bibinfo{pages}{613--626}.
\newblock


\bibitem[Dodhia(2005)]%
        {dodhia2005review}
\bibfield{author}{\bibinfo{person}{Rahul~M Dodhia}.}
  \bibinfo{year}{2005}\natexlab{}.
\newblock \showarticletitle{A review of applied multiple regression/correlation
  analysis for the behavioral sciences}.
\newblock \bibinfo{journal}{\emph{Journal of Educational and Behavioral
  Statistics}} \bibinfo{volume}{30}, \bibinfo{number}{2}
  (\bibinfo{year}{2005}), \bibinfo{pages}{227--229}.
\newblock


\bibitem[Feng et~al\mbox{.}(2022)]%
        {feng-etal-2022-legal}
\bibfield{author}{\bibinfo{person}{Yi Feng}, \bibinfo{person}{Chuanyi Li},
  {and} \bibinfo{person}{Vincent Ng}.} \bibinfo{year}{2022}\natexlab{}.
\newblock \showarticletitle{Legal Judgment Prediction via Event Extraction with
  Constraints}. In \bibinfo{booktitle}{\emph{Proceedings of the 60th Annual
  Meeting of the Association for Computational Linguistics (Volume 1: Long
  Papers)}}. \bibinfo{publisher}{Association for Computational Linguistics},
  \bibinfo{address}{Dublin, Ireland}, \bibinfo{pages}{648--664}.
\newblock
\urldef\tempurl%
\url{https://doi.org/10.18653/v1/2022.acl-long.48}
\showDOI{\tempurl}


\bibitem[Hartford et~al\mbox{.}(2017)]%
        {hartford2017deep}
\bibfield{author}{\bibinfo{person}{Jason Hartford}, \bibinfo{person}{Greg
  Lewis}, \bibinfo{person}{Kevin Leyton-Brown}, {and} \bibinfo{person}{Matt
  Taddy}.} \bibinfo{year}{2017}\natexlab{}.
\newblock \showarticletitle{Deep IV: A flexible approach for counterfactual
  prediction}. In \bibinfo{booktitle}{\emph{International Conference on Machine
  Learning}}. PMLR, \bibinfo{pages}{1414--1423}.
\newblock


\bibitem[Kingma and Adam(2015)]%
        {kingma2015method}
\bibfield{author}{\bibinfo{person}{Diederik~P Kingma} {and}
  \bibinfo{person}{Jimmy~Ba Adam}.} \bibinfo{year}{2015}\natexlab{}.
\newblock \showarticletitle{A Method for Stochastic}.
\newblock \bibinfo{journal}{\emph{Optimization. In, ICLR}}  \bibinfo{volume}{5}
  (\bibinfo{year}{2015}).
\newblock


\bibitem[Kumar et~al\mbox{.}(2011)]%
        {kumar2011similarity}
\bibfield{author}{\bibinfo{person}{Sushanta Kumar}, \bibinfo{person}{P~Krishna
  Reddy}, \bibinfo{person}{V~Balakista Reddy}, {and} \bibinfo{person}{Aditya
  Singh}.} \bibinfo{year}{2011}\natexlab{}.
\newblock \showarticletitle{Similarity analysis of legal judgments}. In
  \bibinfo{booktitle}{\emph{Proceedings of the Fourth Annual ACM Bangalore
  Conference}}. \bibinfo{pages}{1--4}.
\newblock


\bibitem[Liu et~al\mbox{.}({[n.\,d.]})]%
        {liueverything}
\bibfield{author}{\bibinfo{person}{Xiao Liu}, \bibinfo{person}{Da Yin},
  \bibinfo{person}{Yansong Feng}, \bibinfo{person}{Yuting Wu}, {and}
  \bibinfo{person}{Dongyan Zhao}.} \bibinfo{year}{[n.\,d.]}\natexlab{}.
\newblock \showarticletitle{Everything Has a Cause: Leveraging Causal Inference
  in Legal Text Analysis}.
\newblock  (\bibinfo{year}{[n.\,d.]}).
\newblock


\bibitem[Ma et~al\mbox{.}(2021)]%
        {ma2021lecard}
\bibfield{author}{\bibinfo{person}{Yixiao Ma}, \bibinfo{person}{Yunqiu Shao},
  \bibinfo{person}{Yueyue Wu}, \bibinfo{person}{Yiqun Liu},
  \bibinfo{person}{Ruizhe Zhang}, \bibinfo{person}{Min Zhang}, {and}
  \bibinfo{person}{Shaoping Ma}.} \bibinfo{year}{2021}\natexlab{}.
\newblock \showarticletitle{LeCaRD: A Legal Case Retrieval Dataset for Chinese
  Law System}. In \bibinfo{booktitle}{\emph{Proceedings of the 44th
  International ACM SIGIR Conference on Research and Development in Information
  Retrieval}}. \bibinfo{pages}{2342--2348}.
\newblock


\bibitem[Mihalcea and Tarau(2004)]%
        {mihalcea2004textrank}
\bibfield{author}{\bibinfo{person}{Rada Mihalcea} {and} \bibinfo{person}{Paul
  Tarau}.} \bibinfo{year}{2004}\natexlab{}.
\newblock \showarticletitle{Textrank: Bringing order into text}. In
  \bibinfo{booktitle}{\emph{Proceedings of the 2004 conference on empirical
  methods in natural language processing}}. \bibinfo{pages}{404--411}.
\newblock


\bibitem[Minocha et~al\mbox{.}(2015)]%
        {minocha2015finding}
\bibfield{author}{\bibinfo{person}{Akshay Minocha}, \bibinfo{person}{Navjyoti
  Singh}, {and} \bibinfo{person}{Arjit Srivastava}.}
  \bibinfo{year}{2015}\natexlab{}.
\newblock \showarticletitle{Finding relevant indian judgments using dispersion
  of citation network}. In \bibinfo{booktitle}{\emph{Proceedings of the 24th
  International Conference on World Wide Web}}. \bibinfo{pages}{1085--1088}.
\newblock


\bibitem[Niu et~al\mbox{.}(2022)]%
        {niu2022estimation}
\bibfield{author}{\bibinfo{person}{Ziang Niu}, \bibinfo{person}{Yuwen Gu},
  {and} \bibinfo{person}{Wei Li}.} \bibinfo{year}{2022}\natexlab{}.
\newblock \showarticletitle{Estimation and inference for high-dimensional
  nonparametric additive instrumental-variables regression}.
\newblock \bibinfo{journal}{\emph{arXiv e-prints}} (\bibinfo{year}{2022}),
  \bibinfo{pages}{arXiv--2204}.
\newblock


\bibitem[Pearl(2009)]%
        {pearl2009causality}
\bibfield{author}{\bibinfo{person}{Judea Pearl}.}
  \bibinfo{year}{2009}\natexlab{}.
\newblock \bibinfo{booktitle}{\emph{Causality}}.
\newblock \bibinfo{publisher}{Cambridge university press}.
\newblock


\bibitem[Peters et~al\mbox{.}(2017)]%
        {peters2017elements}
\bibfield{author}{\bibinfo{person}{Jonas Peters}, \bibinfo{person}{Dominik
  Janzing}, {and} \bibinfo{person}{Bernhard Sch{\"o}lkopf}.}
  \bibinfo{year}{2017}\natexlab{}.
\newblock \bibinfo{booktitle}{\emph{Elements of causal inference: foundations
  and learning algorithms}}.
\newblock \bibinfo{publisher}{The MIT Press}.
\newblock


\bibitem[Reimers and Gurevych(2019)]%
        {reimers2019sentence}
\bibfield{author}{\bibinfo{person}{Nils Reimers} {and} \bibinfo{person}{Iryna
  Gurevych}.} \bibinfo{year}{2019}\natexlab{}.
\newblock \showarticletitle{Sentence-BERT: Sentence Embeddings using Siamese
  BERT-Networks}. In \bibinfo{booktitle}{\emph{Proceedings of the 2019
  Conference on Empirical Methods in Natural Language Processing and the 9th
  International Joint Conference on Natural Language Processing
  (EMNLP-IJCNLP)}}. \bibinfo{pages}{3982--3992}.
\newblock


\bibitem[Robertson et~al\mbox{.}(2009)]%
        {robertson2009probabilistic}
\bibfield{author}{\bibinfo{person}{Stephen Robertson}, \bibinfo{person}{Hugo
  Zaragoza}, {et~al\mbox{.}}} \bibinfo{year}{2009}\natexlab{}.
\newblock \showarticletitle{The probabilistic relevance framework: BM25 and
  beyond}.
\newblock \bibinfo{journal}{\emph{Foundations and Trends{\textregistered} in
  Information Retrieval}} \bibinfo{volume}{3}, \bibinfo{number}{4}
  (\bibinfo{year}{2009}), \bibinfo{pages}{333--389}.
\newblock


\bibitem[Saravanan et~al\mbox{.}(2009)]%
        {saravanan2009improving}
\bibfield{author}{\bibinfo{person}{Manavalan Saravanan},
  \bibinfo{person}{Balaraman Ravindran}, {and} \bibinfo{person}{Shivani
  Raman}.} \bibinfo{year}{2009}\natexlab{}.
\newblock \showarticletitle{Improving legal information retrieval using an
  ontological framework}.
\newblock \bibinfo{journal}{\emph{Artificial Intelligence and Law}}
  \bibinfo{volume}{17}, \bibinfo{number}{2} (\bibinfo{year}{2009}),
  \bibinfo{pages}{101--124}.
\newblock


\bibitem[Sch{\"o}lkopf et~al\mbox{.}(2021)]%
        {scholkopf2021toward}
\bibfield{author}{\bibinfo{person}{Bernhard Sch{\"o}lkopf},
  \bibinfo{person}{Francesco Locatello}, \bibinfo{person}{Stefan Bauer},
  \bibinfo{person}{Nan~Rosemary Ke}, \bibinfo{person}{Nal Kalchbrenner},
  \bibinfo{person}{Anirudh Goyal}, {and} \bibinfo{person}{Yoshua Bengio}.}
  \bibinfo{year}{2021}\natexlab{}.
\newblock \showarticletitle{Toward causal representation learning}.
\newblock \bibinfo{journal}{\emph{Proc. IEEE}} \bibinfo{volume}{109},
  \bibinfo{number}{5} (\bibinfo{year}{2021}), \bibinfo{pages}{612--634}.
\newblock


\bibitem[Shao et~al\mbox{.}(2020)]%
        {shao2020bert}
\bibfield{author}{\bibinfo{person}{Yunqiu Shao}, \bibinfo{person}{Jiaxin Mao},
  \bibinfo{person}{Yiqun Liu}, \bibinfo{person}{Weizhi Ma},
  \bibinfo{person}{Ken Satoh}, \bibinfo{person}{Min Zhang}, {and}
  \bibinfo{person}{Shaoping Ma}.} \bibinfo{year}{2020}\natexlab{}.
\newblock \showarticletitle{BERT-PLI: Modeling Paragraph-Level Interactions for
  Legal Case Retrieval.}. In \bibinfo{booktitle}{\emph{IJCAI}}.
  \bibinfo{pages}{3501--3507}.
\newblock


\bibitem[Shaver(2005)]%
        {shaver2005testing}
\bibfield{author}{\bibinfo{person}{J~Myles Shaver}.}
  \bibinfo{year}{2005}\natexlab{}.
\newblock \showarticletitle{Testing for mediating variables in management
  research: Concerns, implications, and alternative strategies}.
\newblock \bibinfo{journal}{\emph{Journal of management}} \bibinfo{volume}{31},
  \bibinfo{number}{3} (\bibinfo{year}{2005}), \bibinfo{pages}{330--353}.
\newblock


\bibitem[Si et~al\mbox{.}(2022)]%
        {si2022model}
\bibfield{author}{\bibinfo{person}{Zihua Si}, \bibinfo{person}{Xueran Han},
  \bibinfo{person}{Xiao Zhang}, \bibinfo{person}{Jun Xu}, \bibinfo{person}{Yue
  Yin}, \bibinfo{person}{Yang Song}, {and} \bibinfo{person}{Ji-Rong Wen}.}
  \bibinfo{year}{2022}\natexlab{}.
\newblock \showarticletitle{A Model-Agnostic Causal Learning Framework for
  Recommendation using Search Data}. In \bibinfo{booktitle}{\emph{Proceedings
  of the ACM Web Conference 2022}}. \bibinfo{pages}{224--233}.
\newblock


\bibitem[Si et~al\mbox{.}(2023)]%
        {si_enhancing}
\bibfield{author}{\bibinfo{person}{Zihua Si}, \bibinfo{person}{Zhongxiang Sun},
  \bibinfo{person}{Xiao Zhang}, \bibinfo{person}{Jun Xu}, \bibinfo{person}{Yang
  Song}, \bibinfo{person}{Xiaoxue Zang}, {and} \bibinfo{person}{Ji-Rong Wen}.}
  \bibinfo{year}{2023}\natexlab{}.
\newblock \showarticletitle{Enhancing Recommendation with Search Data in a
  Causal Learning Manner}.
\newblock  \bibinfo{volume}{41}, \bibinfo{number}{4} (\bibinfo{year}{2023}).
\newblock


\bibitem[Stock and Trebbi(2003)]%
        {stock2003retrospectives}
\bibfield{author}{\bibinfo{person}{James~H Stock} {and}
  \bibinfo{person}{Francesco Trebbi}.} \bibinfo{year}{2003}\natexlab{}.
\newblock \showarticletitle{Retrospectives: Who invented instrumental variable
  regression?}
\newblock \bibinfo{journal}{\emph{Journal of Economic Perspectives}}
  \bibinfo{volume}{17}, \bibinfo{number}{3} (\bibinfo{year}{2003}),
  \bibinfo{pages}{177--194}.
\newblock


\bibitem[Sun(2023)]%
        {sun2023short}
\bibfield{author}{\bibinfo{person}{Zhongxiang Sun}.}
  \bibinfo{year}{2023}\natexlab{}.
\newblock \showarticletitle{A Short Survey of Viewing Large Language Models in
  Legal Aspect}.
\newblock \bibinfo{journal}{\emph{arXiv preprint arXiv:2303.09136}}
  (\bibinfo{year}{2023}).
\newblock


\bibitem[Sz{\'e}kely and Rizzo(2014)]%
        {szekely2014partial}
\bibfield{author}{\bibinfo{person}{G{\'a}bor~J Sz{\'e}kely} {and}
  \bibinfo{person}{Maria~L Rizzo}.} \bibinfo{year}{2014}\natexlab{}.
\newblock \showarticletitle{Partial distance correlation with methods for
  dissimilarities}.
\newblock \bibinfo{journal}{\emph{The Annals of Statistics}}
  \bibinfo{volume}{42}, \bibinfo{number}{6} (\bibinfo{year}{2014}),
  \bibinfo{pages}{2382--2412}.
\newblock


\bibitem[Van~der Maaten and Hinton(2008)]%
        {van2008visualizing}
\bibfield{author}{\bibinfo{person}{Laurens Van~der Maaten} {and}
  \bibinfo{person}{Geoffrey Hinton}.} \bibinfo{year}{2008}\natexlab{}.
\newblock \showarticletitle{Visualizing data using t-SNE.}
\newblock \bibinfo{journal}{\emph{Journal of machine learning research}}
  \bibinfo{volume}{9}, \bibinfo{number}{11} (\bibinfo{year}{2008}).
\newblock


\bibitem[VanderWeele(2013)]%
        {vanderweele2013three}
\bibfield{author}{\bibinfo{person}{Tyler~J VanderWeele}.}
  \bibinfo{year}{2013}\natexlab{}.
\newblock \showarticletitle{A three-way decomposition of a total effect into
  direct, indirect, and interactive effects}.
\newblock \bibinfo{journal}{\emph{Epidemiology (Cambridge, Mass.)}}
  \bibinfo{volume}{24}, \bibinfo{number}{2} (\bibinfo{year}{2013}),
  \bibinfo{pages}{224}.
\newblock


\bibitem[Venkatraman et~al\mbox{.}(2016)]%
        {venkatraman2016online}
\bibfield{author}{\bibinfo{person}{Arun Venkatraman}, \bibinfo{person}{Wen
  Sun}, \bibinfo{person}{Martial Hebert}, \bibinfo{person}{J Bagnell}, {and}
  \bibinfo{person}{Byron Boots}.} \bibinfo{year}{2016}\natexlab{}.
\newblock \showarticletitle{Online instrumental variable regression with
  applications to online linear system identification}. In
  \bibinfo{booktitle}{\emph{Proceedings of the AAAI Conference on Artificial
  Intelligence}}, Vol.~\bibinfo{volume}{30}.
\newblock


\bibitem[Wooldridge(2015)]%
        {wooldridge2015introductory}
\bibfield{author}{\bibinfo{person}{Jeffrey~M Wooldridge}.}
  \bibinfo{year}{2015}\natexlab{}.
\newblock \bibinfo{booktitle}{\emph{Introductory econometrics: A modern
  approach}}.
\newblock \bibinfo{publisher}{Cengage learning}.
\newblock


\bibitem[Wu et~al\mbox{.}(2022)]%
        {wu2022treatment}
\bibfield{author}{\bibinfo{person}{Anpeng Wu}, \bibinfo{person}{Kun Kuang},
  \bibinfo{person}{Bo Li}, {and} \bibinfo{person}{Fei Wu}.}
  \bibinfo{year}{2022}\natexlab{}.
\newblock \showarticletitle{Instrumental variable regression with confounder
  balancing}. In \bibinfo{booktitle}{\emph{International Conference on Machine
  Learning}}. PMLR, \bibinfo{pages}{24056--24075}.
\newblock


\bibitem[Xiao et~al\mbox{.}(2021)]%
        {xiao2021lawformer}
\bibfield{author}{\bibinfo{person}{Chaojun Xiao}, \bibinfo{person}{Xueyu Hu},
  \bibinfo{person}{Zhiyuan Liu}, \bibinfo{person}{Cunchao Tu}, {and}
  \bibinfo{person}{Maosong Sun}.} \bibinfo{year}{2021}\natexlab{}.
\newblock \showarticletitle{Lawformer: A pre-trained language model for chinese
  legal long documents}.
\newblock \bibinfo{journal}{\emph{AI Open}}  \bibinfo{volume}{2}
  (\bibinfo{year}{2021}), \bibinfo{pages}{79--84}.
\newblock


\bibitem[Xu et~al\mbox{.}(2020a)]%
        {xu2020learning}
\bibfield{author}{\bibinfo{person}{Liyuan Xu}, \bibinfo{person}{Yutian Chen},
  \bibinfo{person}{Siddarth Srinivasan}, \bibinfo{person}{Nando de Freitas},
  \bibinfo{person}{Arnaud Doucet}, {and} \bibinfo{person}{Arthur Gretton}.}
  \bibinfo{year}{2020}\natexlab{a}.
\newblock \showarticletitle{Learning Deep Features in Instrumental Variable
  Regression}. In \bibinfo{booktitle}{\emph{International Conference on
  Learning Representations}}.
\newblock


\bibitem[Xu et~al\mbox{.}(2020b)]%
        {xu2020distinguish}
\bibfield{author}{\bibinfo{person}{Nuo Xu}, \bibinfo{person}{Pinghui Wang},
  \bibinfo{person}{Long Chen}, \bibinfo{person}{Li Pan},
  \bibinfo{person}{Xiaoyan Wang}, {and} \bibinfo{person}{Junzhou Zhao}.}
  \bibinfo{year}{2020}\natexlab{b}.
\newblock \showarticletitle{Distinguish Confusing Law Articles for Legal
  Judgment Prediction}. In \bibinfo{booktitle}{\emph{Proceedings of the 58th
  Annual Meeting of the Association for Computational Linguistics}}.
  \bibinfo{pages}{3086--3095}.
\newblock


\bibitem[Yang et~al\mbox{.}(2021)]%
        {yang2021causalvae}
\bibfield{author}{\bibinfo{person}{Mengyue Yang}, \bibinfo{person}{Furui Liu},
  \bibinfo{person}{Zhitang Chen}, \bibinfo{person}{Xinwei Shen},
  \bibinfo{person}{Jianye Hao}, {and} \bibinfo{person}{Jun Wang}.}
  \bibinfo{year}{2021}\natexlab{}.
\newblock \showarticletitle{CausalVAE: Disentangled representation learning via
  neural structural causal models}. In \bibinfo{booktitle}{\emph{Proceedings of
  the IEEE/CVF Conference on Computer Vision and Pattern Recognition}}.
  \bibinfo{pages}{9593--9602}.
\newblock


\bibitem[Yu et~al\mbox{.}(2022a)]%
        {yu-etal-2022-optimal}
\bibfield{author}{\bibinfo{person}{Weijie Yu}, \bibinfo{person}{Liang Pang},
  \bibinfo{person}{Jun Xu}, \bibinfo{person}{Bing Su}, \bibinfo{person}{Zhenhua
  Dong}, {and} \bibinfo{person}{Ji-Rong Wen}.}
  \bibinfo{year}{2022}\natexlab{a}.
\newblock \showarticletitle{Optimal Partial Transport Based Sentence Selection
  for Long-form Document Matching}. In \bibinfo{booktitle}{\emph{Proceedings of
  the 29th International Conference on Computational Linguistics}}.
  \bibinfo{publisher}{International Committee on Computational Linguistics},
  \bibinfo{address}{Gyeongju, Republic of Korea}, \bibinfo{pages}{2363--2373}.
\newblock
\urldef\tempurl%
\url{https://aclanthology.org/2022.coling-1.208}
\showURL{%
\tempurl}


\bibitem[Yu et~al\mbox{.}(2022b)]%
        {yu2022explainable}
\bibfield{author}{\bibinfo{person}{Weijie Yu}, \bibinfo{person}{Zhongxiang
  Sun}, \bibinfo{person}{Jun Xu}, \bibinfo{person}{Zhenhua Dong},
  \bibinfo{person}{Xu Chen}, \bibinfo{person}{Hongteng Xu}, {and}
  \bibinfo{person}{Ji-Rong Wen}.} \bibinfo{year}{2022}\natexlab{b}.
\newblock \showarticletitle{Explainable Legal Case Matching via Inverse Optimal
  Transport-based Rationale Extraction}. In
  \bibinfo{booktitle}{\emph{Proceedings of the 45th International ACM SIGIR
  Conference on Research and Development in Information Retrieval}}.
  \bibinfo{pages}{657--668}.
\newblock


\bibitem[Yuan et~al\mbox{.}(2022)]%
        {yuan2022auto}
\bibfield{author}{\bibinfo{person}{Junkun Yuan}, \bibinfo{person}{Anpeng Wu},
  \bibinfo{person}{Kun Kuang}, \bibinfo{person}{Bo Li}, \bibinfo{person}{Runze
  Wu}, \bibinfo{person}{Fei Wu}, {and} \bibinfo{person}{Lanfen Lin}.}
  \bibinfo{year}{2022}\natexlab{}.
\newblock \showarticletitle{Auto IV: Counterfactual Prediction via Automatic
  Instrumental Variable Decomposition}.
\newblock \bibinfo{journal}{\emph{ACM Transactions on Knowledge Discovery from
  Data (TKDD)}} \bibinfo{volume}{16}, \bibinfo{number}{4}
  (\bibinfo{year}{2022}), \bibinfo{pages}{1--20}.
\newblock


\bibitem[Zeng et~al\mbox{.}(2005)]%
        {zeng2005knowledge}
\bibfield{author}{\bibinfo{person}{Yiming Zeng}, \bibinfo{person}{Ruili Wang},
  \bibinfo{person}{John Zeleznikow}, {and} \bibinfo{person}{Elizabeth Kemp}.}
  \bibinfo{year}{2005}\natexlab{}.
\newblock \showarticletitle{Knowledge representation for the intelligent legal
  case retrieval}. In \bibinfo{booktitle}{\emph{International Conference on
  Knowledge-Based and Intelligent Information and Engineering Systems}}.
  Springer, \bibinfo{pages}{339--345}.
\newblock


\bibitem[Zhong et~al\mbox{.}(2018)]%
        {zhong2018legal}
\bibfield{author}{\bibinfo{person}{Haoxi Zhong}, \bibinfo{person}{Zhipeng Guo},
  \bibinfo{person}{Cunchao Tu}, \bibinfo{person}{Chaojun Xiao},
  \bibinfo{person}{Zhiyuan Liu}, {and} \bibinfo{person}{Maosong Sun}.}
  \bibinfo{year}{2018}\natexlab{}.
\newblock \showarticletitle{Legal judgment prediction via topological
  learning}. In \bibinfo{booktitle}{\emph{Proceedings of the 2018 Conference on
  Empirical Methods in Natural Language Processing}}.
  \bibinfo{pages}{3540--3549}.
\newblock


\end{thebibliography}

\end{document}